\documentclass[conference]{IEEEtran}

\usepackage{cite}
\usepackage{times}
\usepackage{dsfont}

\usepackage{graphicx}
\usepackage{graphics}
\usepackage{subfigure}
\usepackage{epsfig}
\usepackage{latexsym}
\usepackage{amsfonts}
\usepackage{amssymb}
\usepackage{paralist}
\usepackage{comment}
\usepackage{xspace}
\usepackage{mathrsfs}
\usepackage{color}

\usepackage{url,epsfig,array}
\usepackage{leftidx}
\usepackage{amsmath}
\usepackage{url}

\usepackage{paralist}
\usepackage[lined,linesnumbered]{algorithm2e}

\def\ie{\textit{i.e.}\xspace}

\def\eg{\textit{e.g.}\xspace}

\newcommand{\equref}[1]{{Eq.~(\ref{#1})}}

\begin{document}

\title{SmartLoc: Sensing Landmarks Silently for Smartphone Based Metropolitan Localization}
\author{
 \authorblockN{Cheng Bo\authorrefmark{1},
 Xiang-Yang Li\authorrefmark{1}
 Taeho Jung\authorrefmark{1},
 Xufei Mao\authorrefmark{2}}
 \authorblockA{
\authorrefmark{1} Department of Computer Science, Illinois Institute of Technology, USA\\
 \authorrefmark{2} School of Software and TNLIST, Tsinghua University, Beijing, China\\
 Email: cbo@hawk.iit.edu, xli@cs.iit.edu, tjung@hawk.iit.edu, xufei.mao@gmail.com}}

\maketitle

\begin{abstract}
%GPS is the most widely used solution for localization and navigation.
%However, it often works poorly in areas with bad GPS signals, e.g., urban areas
% with tall buildings.
We present \emph{SmartLoc}, a localization system to estimate the location and the
 traveling distance by leveraging the lower-power inertial sensors
 embedded in smartphones as a supplementary to GPS.
To minimize the negative impact of sensor noises,
 \emph{SmartLoc} exploits the intermittent strong GPS signals and uses
 the linear regression to
 build a prediction model which is based on the trace estimated from inertial
 sensors and the one computed from the GPS.
Furthermore, we utilize landmarks (\eg, bridge, traffic lights)
 detected automatically and special driving patterns (\eg, turning, uphill, and downhill) from
 inertial sensory data to improve the  localization accuracy when
 the GPS signal is weak.
Our evaluations of \emph{SmartLoc} in the city
 demonstrates its technique viability and significant localization accuracy
 improvement compared with GPS and other approaches: the error is
 approximately $20$m for $90\%$ of time while the known mean error of GPS is $42.22$m.
%\emph{SmartLoc} system can also reduce the energy
% consumption by at least $10\%$  for localization in other scenarios
% by carefully turning on GPS periodically without sacrificing
% localization accuracy.

\end{abstract}


\section{Introduction}
\label{sec:intro}
Localization have attracted significant attentions in the past few years,
 and numerous techniques have been proposed to achieve high accuracy localization.
In outdoor scenarios, GPS and its variants
 are the most common technologies to provide accurate position for various
 applications~\cite{cheng2005accuracy}.
However, problems caused by weak/none GPS signal in cities often
 lead to a pretty bad user experience.
% or indoor environment
% have promoted the idea of war-driving and created a number of
% WiFi/GSM fingerprint based localization strategies, like
% Skyhook~\cite{skyhook}, which are however computationally intensive.
For instance, we conduct comprehensive experiments in downtown Chicago, IL USA,
 to evaluate the performance of GPS positioning.
Based on the experiment results, we observe that the GPS signals are very weak and
 unstable in some roads due to highrises, or even blocked completely in
 some complicated road structures, such as tunnels and underground.
In addition, the largest location error we collected is over $100$m on the ground, and nearly $400$m in the underground
 segments (see Fig.~\ref{fig:pex} for more details).
Thus,  improving the location accuracy is imperative when the GPS signal
 is weak.

In this work, we propose \emph{SmartLoc}, a localization method which improves
 the localization accuracy in metropolises by leveraging
 embedded inertial sensors in smartphones to help improve the driving patterns according
 to various of road conditions.
Exploiting the data collected from these inertial sensors has
 been used in the literature to address a number of challenging and
 interesting tasks, \eg, indoor localization
 \cite{constandache2010towards,wang2012no,yang2012locating,liu2012push},
 road condition monitoring \cite{mohan2008nericell,eriksson2008pothole},
 property tracking~\cite{guha2010autowitness}, and
 outdoor localization \cite{guha2010autowitness,hwanggps,paek2010energy}.
Note that some applications exploits accelerometer to measure walking speed and distance
 of pedestrian~\cite{chen2009integrated,constandache2010towards,retscher2006intelligent,wang2012no}
 and exploit compass to estimate the direction so as to estimate the location.
However, providing realtime localization of moving cars in metropolises
 is far more challenging as such activity does not have a cyclic pattern in sensory data.

To address these challenges, during the dead reckoning process for calculating
 the current position of a car, we propose a dynamic trajectory model to estimate
 the driving speed and velocity based on current road condition, so that the impact of
 inherent noise and accumulated error could be reduced to a large extent.
We also design a calibration strategy based on
 road infrastructures (\eg, bridge, traffic
  lights, uphill, and downhill) and driving status (\eg, turns,
  stops), which are inferred from the sensory data.
Our extensive evaluations indicate that leveraging inertial sensors could
 accurately identify the special road infrastructures using either
 fingerprint based approaches or pattern-matching technique.
%And \emph{SmartLoc} can provide a much accurate position based on both the
% driving status and matched road infrastructures, which are generated by exploiting
% the current coarse-grained estimation of location from  dead-reckoning to
% confine the search space.
%Our extensive evaluations show that using inertial sensors can
% accurately identify special road infrastructures using either
% fingerprint based approaches or pattern-matching technique.
%\emph{SmartLoc} also exploits the current coarse-grained estimation of
% location from  dead-reckoning to confine the search space which often
% has only one candidate matching left.
%Based on the matched road infrastructures or driving status, SmartLoc
% can provide a much accurate localization.
%Our  evaluations show that turning (left or right), uphill
% and downhill provide localization accuracy within a few meters while
% traffic lights and stop signs provide less accurate localization: the curve of `\emph{SmartLoc}' is closer to the ideal diagonal line than the one of `Inertial Sensor \& Traffic Light' in Figure~\ref{fig:loc_gps_est}.

We implement \emph{SmartLoc} on Android, and evaluate the localization performance in
 both downtown Chicago and highway.
Our extensive test results in the majority of blocks
 in Chicago indicate that \emph{SmartLoc} improves the location accuracy
 significantly:
1) the mean localization error in each time slot is $11.65$m;
2) according to the proportion of "good" road segments, the
 average localization error is less than $20$m such that the
 localization accuracy is increased from
  $\le 50\%$ (by purely using GPS) to $\ge 90\%$ using \emph{SmartLoc} in
   downtown areas.
When testing \emph{SmartLoc} on highway, the localization error is at most
 $12$m for $95\%$ of the time.
In comparison the state-of-the-art localization scheme for moving
 vehicles,  AutoWitness~\cite{guha2010autowitness}, only produces the
 error of distance estimation less than $10\%$ for most of the
 cases,  which could be large when the estimated distance is long
 (\eg, $10\%$ of the 2 miles driving is $320$m).
Our results also imply that \emph{SmartLoc} can save the energy consumption
 by switching on/off  the GPS periodically.

The main contributions of this work are summarized as follows:
\begin{compactenum}
\item We propose a self-learning driving model to reduce the speed and
 trajectory distance estimation error brought by both the inherent noise
 and dead-reckoning.
\item In a real scenario, when both the traffic condition and road infrastructures are
 complex and unpredictable, which hinder the trajectory estimation accurately,
 \emph{SmartLoc} could adjust the self-learning driving model to calculate
 the best parameters to match the current driving condition.
\item  Although self-calibration is a reliable approach to elevate the accuracy in
 localization, it is still difficult to calibrate the location in metropolises with
 weak GPS signal.
 \emph{SmartLoc} also exploits the current coarse-grained estimation of location
 to confine the search space, so that a much more accurate localization could be
 achieved through matching the road infrastructures and driving status.
\end{compactenum}

The rest of paper is organized as follows.
We first review the state-of-art localization techniques in Section~\ref{sec:rw}.
We show our measurement results and observations with respect to the
 GPS accuracy in Section~\ref{sec:preliminary}.
We present the overview of \emph{SmartLoc} in Section~\ref{sec:overview}, following which
 novel calibration techniques of \emph{SmartLoc} are presented one by one in
 Section~\ref{sec:regression} and Section~\ref{sec:landmarks}.
We report our detailed real-world experiment results in
 Section~\ref{sec:experiment},  and conclude the work in Section~\ref{sec:conclusion}.

\section{Related Works}
\label{sec:rw}
Our work involves in a number of techniques, in this section, we mainly
 focus on the work related to wireless localization and dead-reckoning~\cite{levi1996dead}.

\subsection{Localization Techniques}
GPS~\cite{liu2012energy}, being the most popular outdoor localization
 system, has been widely used to provide localization and navigation services to users
 such that numerous techniques have been proposed in the literature to improve
 the GPS localization accuracy, like A-GPS, D-GPS~\cite{parkinson1996differential},
 WAAS~\cite{enge1996wide}, etc.
Recently, WiFi signal \cite{cheng2005accuracy,skyhook} and cellular signal
 \cite{varshavsky2005gsm,chen2006practical} have been used to find the
 locations as well.
However, the median error for downtown environment based on cellular signal
 reaches $100$m at worst\cite{chen2006practical}, and WiFi based solutions rely
 on nearby WiFi APs' locations.
Unfortunately, these GPS-based or WiFi-based solutions are inapplicable for navigation in metropolises because
 of many critical road infrastructures, such as under
 ground roads and multilayered roads where the GPS signal is often lost,
  and there are no WiFi access points at all.
Some GSM-based localization methods, like \cite{varshavsky2005gsm, mohan2008nericell},
 are widely available.
However, their accuracy is low (up to hundreds of meters)
 with the assumption that the exact positions of cellular towers
 should be known \emph{in priori}.

Work PlaceLab~\cite{cheng2005accuracy}, and
 ActiveCampus~\cite{griswold2004activecampus} make full use of WiFi and
 GSM signals for location at outdoor environment.
The former creates a map by war-driving
 a region and maps both APs and cell tower's signals to the
 wireless map.
The latter is quite similar except it assumes the APs'
 location is known \emph{in priori}.
Taking advantages of two aforementioned systems,
 CompAcc~\cite{constandache2010towards} uses dead-reckoning combined
 with AGPS to further calibrate localization results rather than utilizing preliminary war-driving.
Unfortunately, all these systems need time-consuming calibration, and are not
 suitable for large scale area.
Another work Skyhook~\cite{skyhook} supply high accuracy
 location services with cost of hiring over $500$ drivers to create the WiFi/GSM map in
 certain region.

Several promising techniques such as  crowdsourcing are
 introduced in localization recently, such as Zee~\cite{rai2012zee},
 which also uses inertial sensors to track users' movement.

\subsection{Dead-Reckoning}
Recently, dead-reckoning strategies using internal sensors to estimate
 motion activities have attracted many research interests.
Strapdown Inertial Navigation System
(SINS)~\cite{titterton2004strapdown} and
Pedometer System~\cite{jirawimut2003method} use MEMS to estimate the
 moving speed and trace. The key issue is to deal with the noise of
 internal sensors and accumulated errors, which sometimes grow
 cubically~\cite{woodman2008pedestrian}.
Personal Dead-reckoning (PDR)
 system~\cite{ojeda2007non}  uses ``Zero Velocity Update''  to calibrate
 the drift.
The majority of the dead-reckoning studies focus on walking
 estimation, such as UnLoc~\cite{wang2012no}, and
 CompAcc~\cite{constandache2010towards}.
Their main idea is to use accelerometer sensors
 to estimate the number of walking steps, and then measure the walking distance.
AutoWitness~\cite{guha2010autowitness} is the system with an
 embedded wireless tag integrated with vibration,
 accelerometer, and gyroscope sensors.
The tag is attached to a vehicle, and accelerometer and gyroscope sensors
 are used to track the moving trace.

\subsection{Road, Map and Traffic}
Smartphones are used to analyze traffic patterns to provide
 better navigation system in vehicle.
CTrack~\cite{thiagarajan2011accurate} and VTrack~\cite{thiagarajan2009vtrack}
 are two systems which process error-prone positioning systems to estimate the trajectories.
These two system match a sequence of observations on the transitions between locations,
 while the former adopt fingerprints and the latter mainly utilizes HMM.
SmartRoad~\cite{hu2013smartroad} detects and identifies traffic lights
 and stop signs through crowd-sensing strategies.
Some research propose map matching algorithms based on Kalman Filter~\cite{obradovic2006fusion} or HMM~\cite{newson2009hidden}.
However, such approaches cannot guarantee accuracy.
IVMM~\cite{yuan2010interactive} is then proposed to increase the accuracy.

\section{GPS Positioning in Downtown}
\label{sec:preliminary}
\subsection{Measurement in Downtown Chicago }
\begin{figure}[h]
\centering
\subfigure[Testing area in Chicago\label{fig:loop}]
{\includegraphics[width=1.6in]{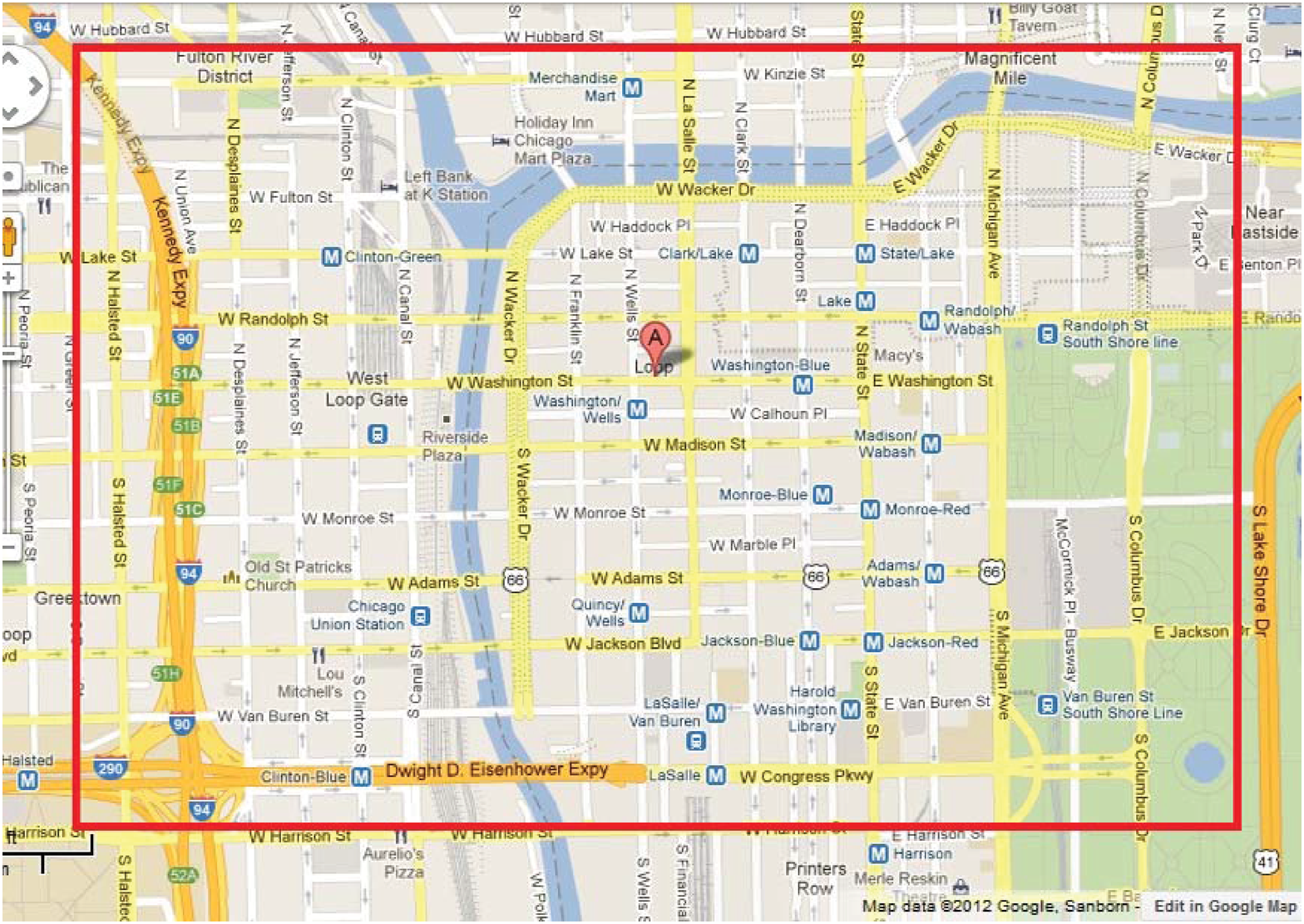}}
\subfigure[Proportion of Error\label{fig:error_pro}]
{\includegraphics[width=1.7in]{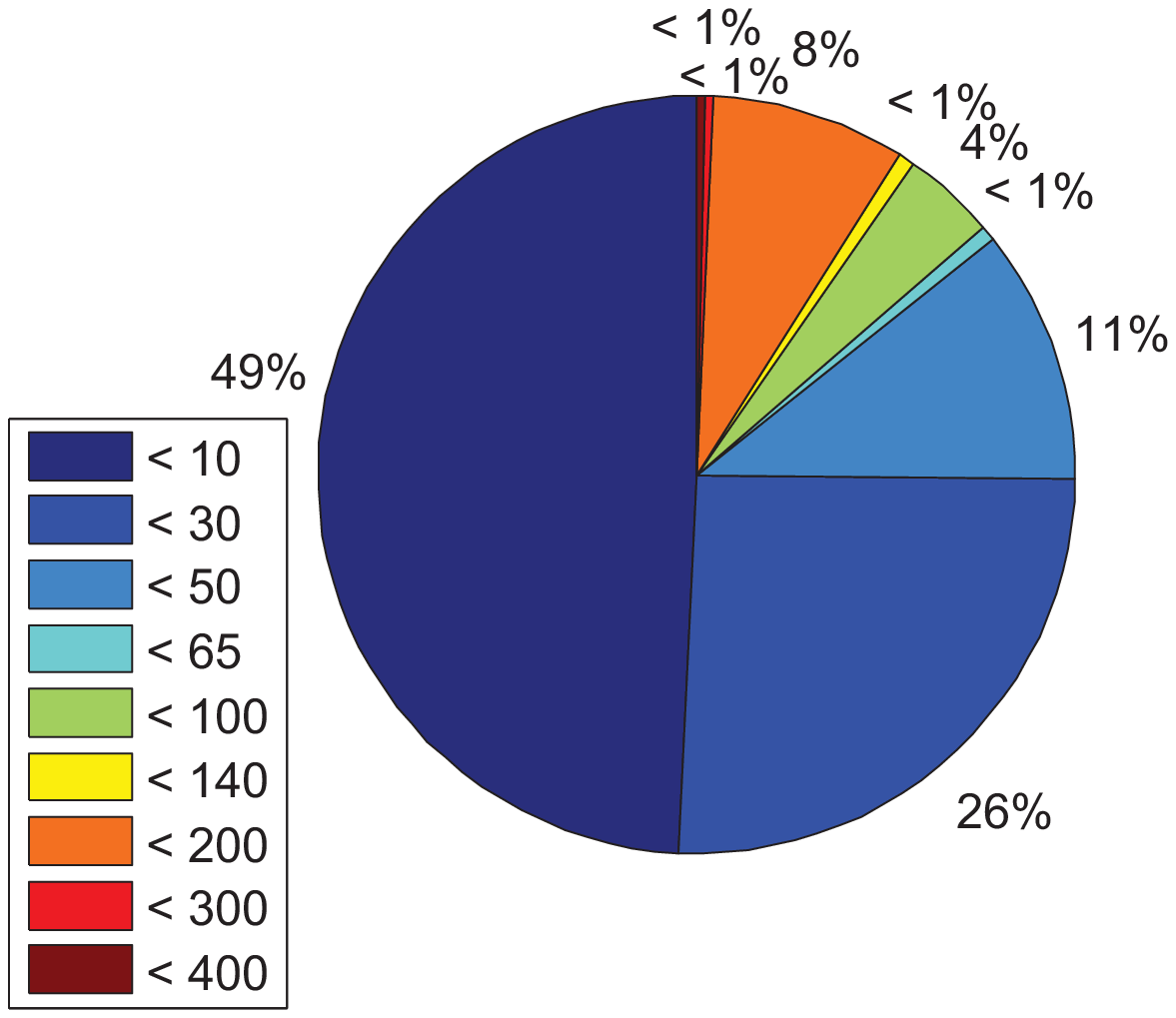}}
\caption{GPS localization accuracy in Chicago.}
\label{fig:pex}
\end{figure}

To study how bad the GPS location accuracy could be,
 we first conduct a comprehensive measurements of GPS accuracies
 within some area in Downtown Chicago, red rectangle shown in
 Figure~\ref{fig:loop}.
We drive through every road in the area while recording location information
 in a real-time manner.
In order to remove the time dependent GPS location errors, we conduct
 independent measurements at three different times, and report the
 results by average.
We find that in the test area, the largest location error
  reaches $400$m, and the distance of the longest road
  segment between two GPS locations with reasonable accuracies ($\le
  30m$) is about $1$km.

\subsection{Original Location Results}
Unfortunately, the location accuracy is not as high as expected according to
 our measurement results.
For instance, the localization results have averagely $42.22$m errors
 and the largest error reaches $400$m, which is nearly the length of three blocks in
 downtown area.
We further plot the localization accuracy information of downtown Chicago based
 on the measurement results in Figure \ref{fig:error_pro}.
Clearly, only about half of the sampling points endure
 the error of less than $20$m, over one quarter of the locations have an error
 of about $50$m while the rest quarter has an error larger than $50$m.

\begin{figure}[h]
\centering
\subfigure[Bad road segments.\label{fig:seg_no_gps}]{\includegraphics[width=1.6in]{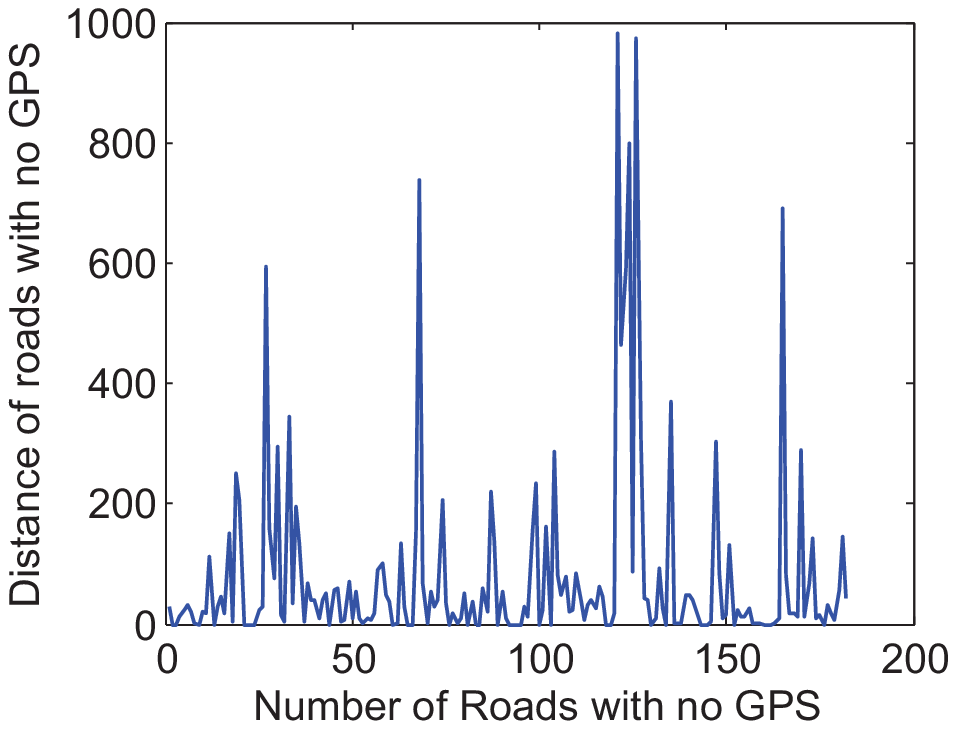}}
\subfigure[Number of bad segments.\label{fig:num_no_30}]{\includegraphics[width=1.6in]{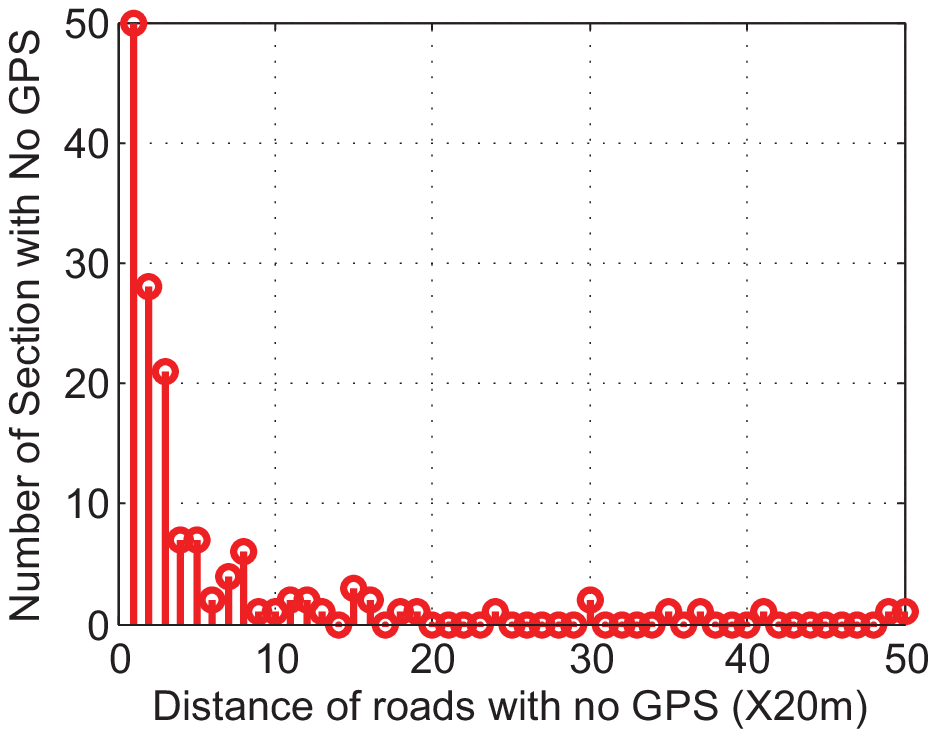}}
\vspace{-0.1in}
\caption{Road segments with poor GPS\label{fig:no_gps}}
\end{figure}

We assume that the largest location error a user could accept inside a city
 should be less than $30$m, which is less than a quarter of one block.
From now on, we consider the positions with GPS location error less than $30$m as
  the locations with \textbf{good GPS signals},
  and the rest as the locations with \textbf{bad GPS signals}.
Since longer segments of roads with bad GPS signals are prone to
  leading to wrong instruction for turning or stopping in a navigation system.
We calculate the distance of road segments with bad GPS signal in the experiment area,
 and present the results in Figure~\ref{fig:no_gps}.
In Figure~\ref{fig:seg_no_gps}, we numbered each segment of road with bad GPS signal in
 X axis, and plot the length of all $182$ segments of road, which indicates that the
 longest length reaches almost one kilometer.
Meanwhile, the Figure \ref{fig:num_no_30} illustrates the distribution of each segment of
 roads.
We notice that the average length of these bad segments of road is approximately
 $200$m, and those with over $400$m locate in the center of downtown, which may confuse
 drivers most.

\vspace{-0.1in}

\section{System Overview}
\label{sec:overview}
\subsection{Main Idea}
The objective of \emph{SmartLoc}is to use
 inertial sensors in smartphones to lively estimate the
 locations based on the trajectory and orientation through self-learnt dynamic
 model with high accuracy but low energy consumption.
Remarkably, we not only address the inaccuracy caused by the complex infrastructures
 in downtown area, but also exploit them to improve the localization accuracy.

In trajectory measuring, traditional methods leveraging inertial
 sensors introduce large inherent errors, which leads to poor traveling distance
 and speed estimation.
Besides the mechanism noise, such errors also come from the process of extracting and transforming
 linear acceleration in Earth Frame Coordinate and orientation estimation.
Although Extended Kalman Filter could be adopted to reduce such coarse noise to some extent, trajectory
 calculating error still cannot be neglected.
%Our main idea is to propose a newly developed localization model to compensate the inaccurate localization in downtown area. SmartLoc collects both GPS information and sensor data, and conducts
% coarse noise reduction by computing the moving average with a certain sliding window.
%Since the large inherent error from sensors will lead to poor distance estimation,
% our system uses Extended Kalman Filter (EKF) to estimate the orientation before
% extracting the linear acceleration of the vehicle.
In the following stage, we propose a self-learning predictive regression model
 to estimate the moving distance based on the extracted acceleration,
 in which the accumulated errors are minimized in the following way.
\emph{SmartLoc} switches to the training process to train the predictive model
 when GPS signal is good.
When GPS signals are unreliable, it uses the trained model to predict the moving trajectory
 of the vehicle.
Due to the complex road conditions and unpredictable driving activities,
 the training process should be updated periodically in our model.
In addition, \emph{SmartLoc} also detects the \textit{landmarks} by finding special patterns
 from sensory data when the car goes through bridges, tunnels, traffic lights or turning points,
 while calibrating the estimation accordingly.

\subsection{Challenges}
Many technical issues should be addressed here.
The first issue is how to design an improved self-learning trajectory estimation
 model according to current driving conditions since naive methods using Newton's Law
 accumulate the noises (e.g., when we double the integral on acceleration results in the displacement,
 the noises are doubly accumulated as well).
The second issue is how to recognize the landmarks, which will be further used to improve
 the localization accuracy in our system.
The last but not least challenge comes from the fact that even if some special landmarks are recognized,
 traffic conditions also affect the localization estimation accuracy, e.g., the unpredictable
 length of waiting queues in front of traffic lights.
 challenges in detail.

\section{Trajectory Calculation}
\label{sec:regression}
\subsection{Background}
Although accelerometer, gyroscope and magnetometer sensors could provide sensory
 data to reflect the motion conditions, the intrinsic noise could make the
 naive distance estimation based on Newton's Law unavailable because the error
 will be accumulated.
%Accelerometer, gyroscope and magnetometer are three inertial sensors
% which are integrated in most modern smartphones to monitor the motion
% behaviors of carrier.
%Although these sensors provide acceleration, angular velocity and the magnet strength,
% the intrinsic mechanical noise of the sensors make the naive distance estimation
% based on Newton's Law impossible due to the accumulated noises.

Since drivers have been used to mount their smartphones on the windshield
 as navigators, and the orientation of the smartphone changes irregularly
 due to driving direction changing and vehicle vibration,
%As basic component for localization scheme, and it has more significant
% role in because our system detects the driving condition (e.g., turning, downhill, uphill) and
% the landmarks (e.g., tunnel, bridge) based on the orientations.
 we build an estimation model through gyroscope-based Extend Kalman Filter
 to decrease the orientation error, and extract linear acceleration
 in the coordinate of Earth.

\subsection{Self-learning Predictive Model}
%Although the distance estimation could be achieved after applying a double
% integration on the acceleration based on Newton's Law,
% the noises from the accelerometer are accumulated during the integral,
% and the error gets enormously huge in just several minutes.
We observe from our preliminary experiments (Section \ref{sec:preliminary}) that
 the majority of the road segments with bad GPS  signals (error $\geq$ 30m) are
 usually shorter than $400$m, which takes about $20-30$ seconds to drive through
 in a normal condition.
On the one hand, such distance is long enough to navigate drivers to wrong places,
 on the other hand it is short enough to endure the errors to some extent.
Therefore, we propose the following predictive dynamic trajectory estimating model which
 adaptively calibrates itself using GPS signals and dead-reckoning.

\textbf{Velocity Estimator:}
Because of the inherent noises and measurement errors, the traditional velocity
 estimation model is no longer reliable.
In this case, we denote the velocity $V_i$ at the end of a timeslot $i$ as
\begin{equation}
V_i = V_{i-1} + \beta\cdot{a_i}\cdot{\Delta{t}} + \mu
\label{eq:v_init}
\end{equation}
where $\beta$ is the parameter to be learned and adjusted in real
time,  $a_i$ is the average measured acceleration during
the timeslot $i$, and $\mu$ is the noise.

When GPS signals are strong, both $V_i$ and $V_{i-1}$ could be achieved
 from the GPS directly, and the mean linear acceleration $a_i$ is extracted from the accelerometer.
Then we regress the model to find the best $\beta$, and calculate the noise $\mu$
 hiding behind.
When the localization through GPS is unreliable, we use the trained model
 proposed to predict the velocity $V_i$.
%, where $\mu$ is sampled from a normal distribution with mean 0.\medskip

\textbf{Distance Estimator:}
For general cases, the trajectory distance gathered from GPS indicates the
 distance with some error.
Therefore, letting $G(\Delta{t}_i)$ be the distance during a timeslot $i$
  read from GPS, which could be presented as:
\begin{displaymath}
G(\Delta{t}_i) = {\lambda}_1\cdot{V_{i-1}}\cdot{\Delta{t}} +
\frac{1}{2}\cdot{\widehat{a_i}}\cdot{\Delta{t}^2} + \eta
\end{displaymath}
where $\widehat{a_i}$ is the actual acceleration in the time slot $i$.
Here $\lambda_1$ is multiplied to reflect the error in the estimated
 speed $V_{i-1}$ for the time slot $i-1$.
Since the known measured acceleration $a_i$ contains both inherent
 noise and measurement errors, by assuming that these error follows normal
 distribution,
 we define the measured acceleration as:
 $$a_i = (1 + \varepsilon)\widehat{a_i} + \delta,$$
%\begin{displaymath}
%a_i = (1 + \varepsilon)\widehat{a_i} + \delta
%\end{displaymath}
where $\widehat{a_i}$ is considered as the true acceleration which cannot be obtained.
Then, we use the following formula to estimate the distance
$G(\Delta t_i)$:

{\small\begin{equation}
G(\Delta t_i)={\lambda}_1\cdot{V_{i-1}}\cdot{\Delta{t}} + {\lambda}_2\frac{1}{2}\cdot{a_i}\cdot{\Delta{t}^2}
                + {\lambda}_3\cdot{\Delta{t}^2} + {\lambda}_4\cdot{\Delta{t}} + \eta
\label{eq:dis_k_gps}
\end{equation}}
where $\lambda_1, \cdots, \lambda_4$ are parameters to be learned by our regression model.
%Here ${\lambda}_2=\frac{1}{1 + \varepsilon}$,
%${\lambda}_3=-\frac{\delta}{2\cdot{(1 + \varepsilon})}$.
When GPS signals are strong (GPS error is $\le 20$m), based on the $V_{i-1}$, $a_i$ is computed
 using the sensory data and the distance from GPS, we train our model using \equref{eq:dis_k_gps},
 which is in turn used to predict the
 distance $G(\Delta{t}_i)$ in the time slot $i$ when GPS signals are
 bad.
From the predicted trajectory distance $G(\Delta t_i)$, the location at the timeslot $i$ could be
 estimated based on the obtained location, distance and orientation.

However, since the location errors from GPS changes in both spacial and temporal dimensions,
 it is difficult to estimate the times and places at which GPS signals become weak.
In addition, driving in downtown area face unpredictable traffic conditions and road
 infrastructures, which affects the parameters learnt from the previous model.
Therefore, we propose a more flexible dynamic adjusting strategy to update the parameters
 to match the current driving status.
In our strategy, we calculate the parameters in predictive dynamic trajectory estimating model
 only based on the latest driving data.
We allocate a small buffer to save the latest driving informations.
When the protocol is still in the learning process, the model will replace the oldest data with
 latest informations in order to update the model parameters.
Based on our evaluation, the estimation accuracy in trajectory distance reduces to a large extent.

\subsection{Movement Detection}
\begin{figure*}
%  \begin{minipage}[t]{0.66\linewidth}
    \centering
        \subfigure[Acceleration\label{fig:acc_motion}]{\includegraphics[scale = 0.25]{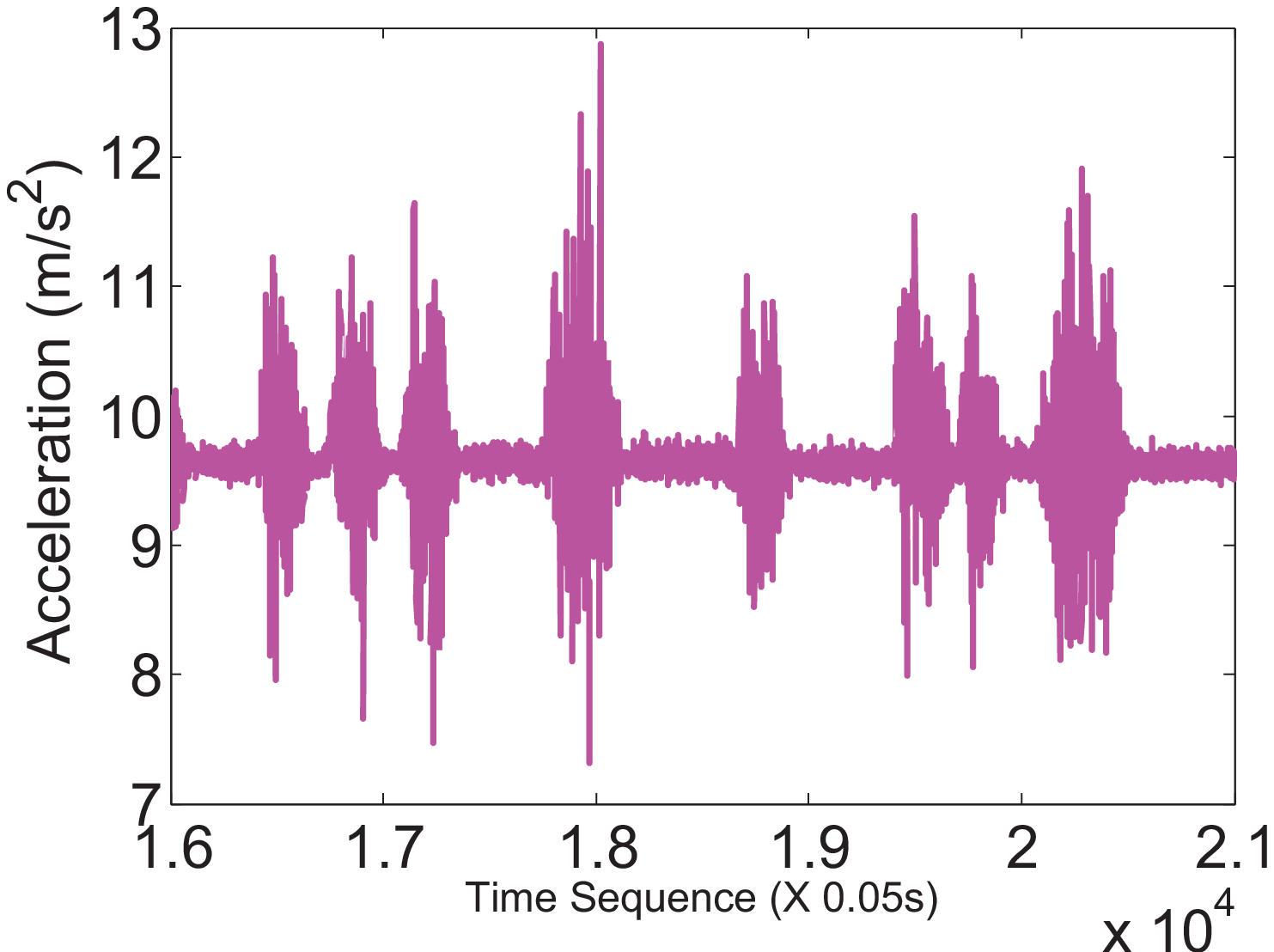}}
        \subfigure[BFC/EFC angle\label{fig:gyeo_motion}]{\includegraphics[scale = 0.25]{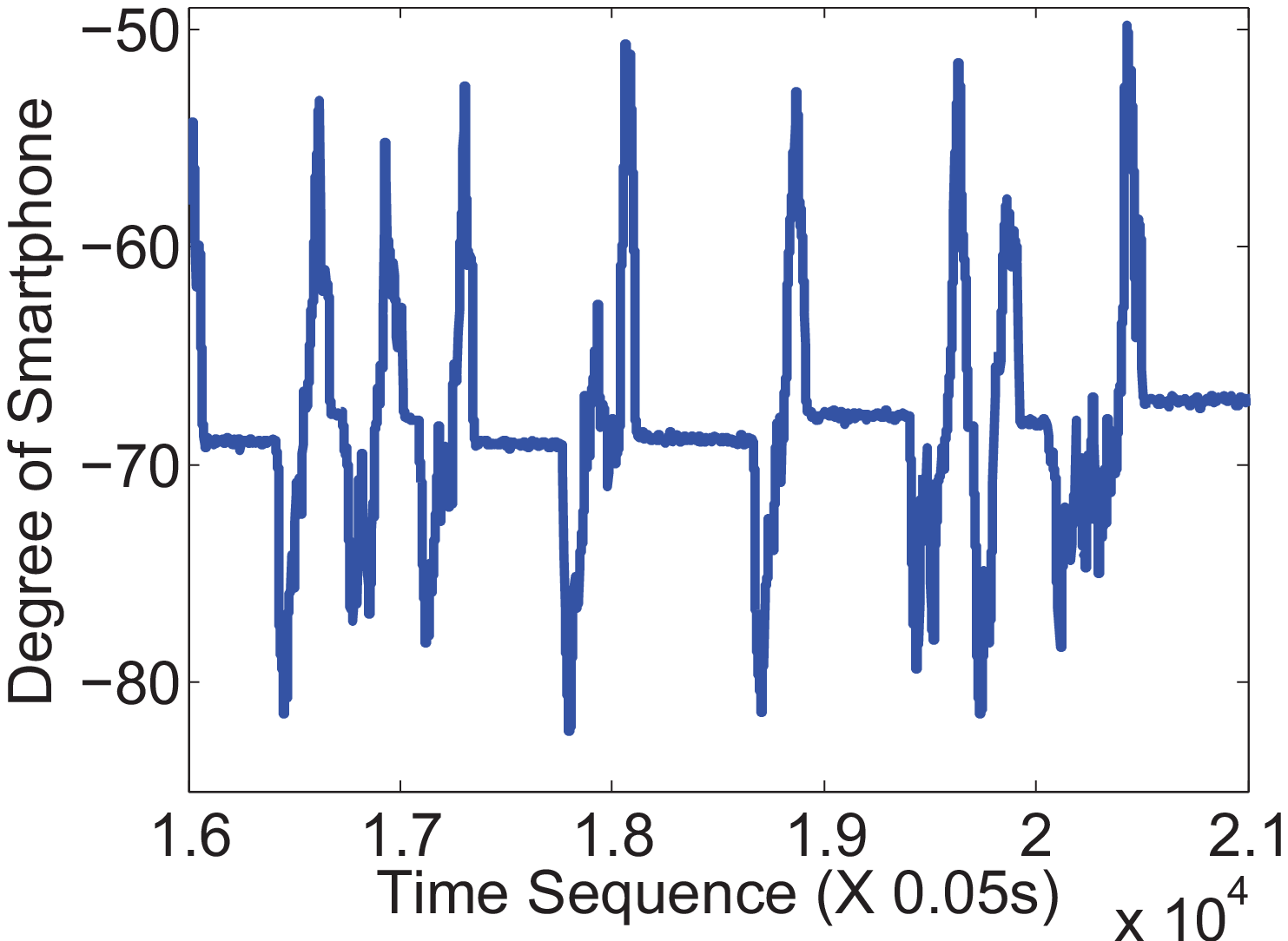}}
        \subfigure[Acceleration in Cruise\label{fig:cruise_acc}]{\includegraphics[scale = 0.25]{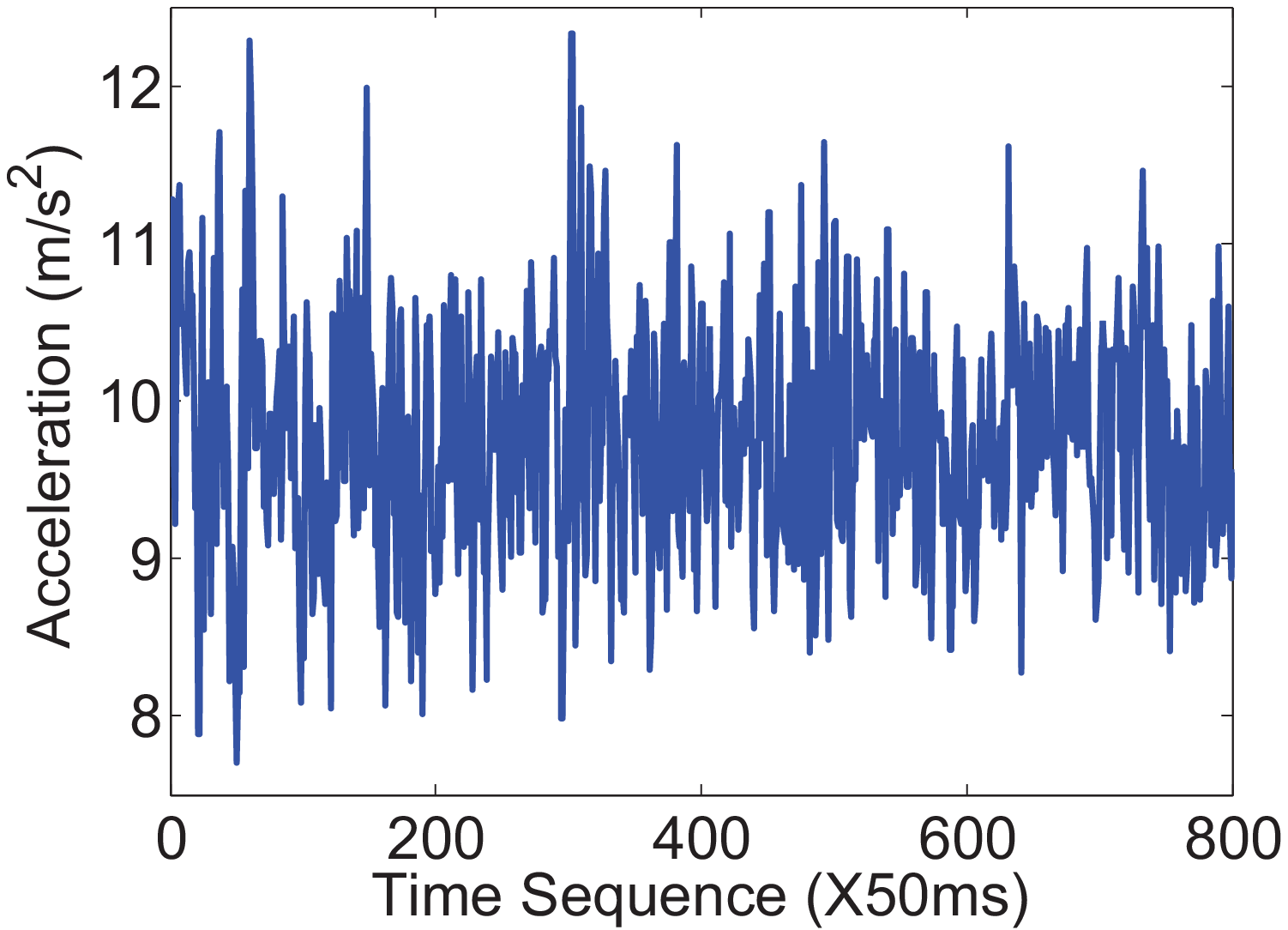}}
        \subfigure[BFC/EFC angle in Cruise\label{fig:cruise_gyro}]{\includegraphics[scale = 0.25]{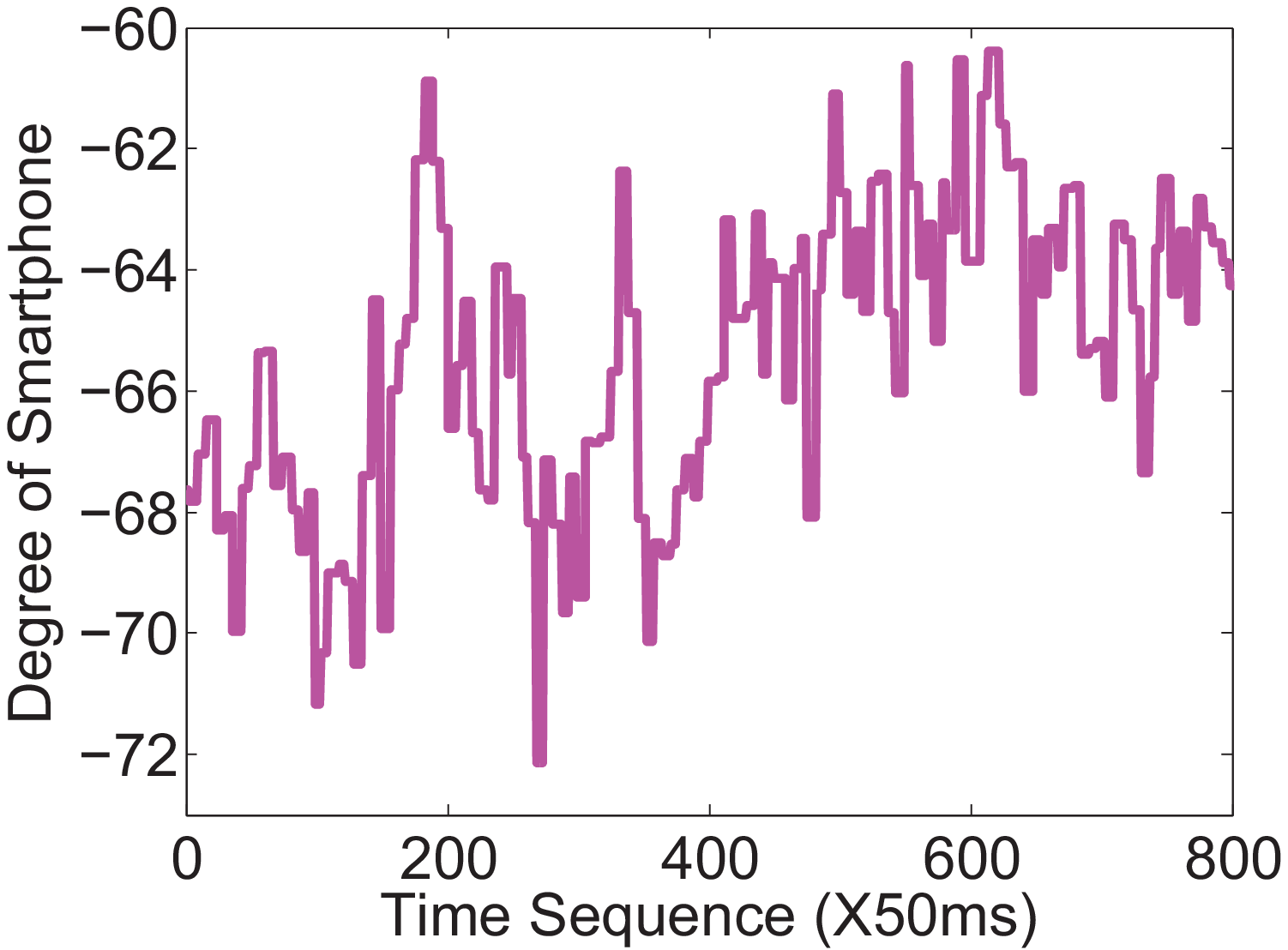}}
        \vspace{-0.12in}
        \caption{The Acceleration and Angle while driving in city or
          Cruise model.}
          \vspace{-0.2in}
\label{fig:motion_detection}
\end{figure*}
%One important problem should be tackled before using our predictive regression model.
Remembering that the speed estimator calculates speeds based on the accelerometer,
 and the speed contains noises accumulated from the integral.
Therefore, even if the vehicle stops, the estimated speed is highly likely to be non-zero,
 which may lead to a huge error in the final prediction.
Hence, determining whether the vehicle is moving or halting  could further reduce the
 negative impact of the mechanical noises.
In addition, movement detection is also the key to the process of landmark
 calibration, which adjusts the location when the vehicle stops in front of traffic lights or stop signs.

During our preliminary experiments, we find that the movement can be reflected
 precisely from both accelerometer and gyroscope sensors, as shown in Figure~\ref{fig:motion_detection}.
The acceleration fluctuates frequently when the vehicle is in motion,
 even in cruise mode, and remains relatively stable when it stops
 (Figure~\ref{fig:acc_motion} and~\ref{fig:cruise_acc}).
The same situation occurs in the gyroscope (Figure~\ref{fig:gyeo_motion}
 and \ref{fig:cruise_gyro}).
Although the smartphone is usually mounted to the windshield,
 due to the inertia while driving, especially speeding up or brake,
 the gyroscope could still sense small rotation changes.
For all the cases, we calculate the variance for readings from both sensors,
 and we find that the largest differences between two vehicles is stopping
 and moving.
For the acceleration, the variance in motion is approximately $60$ times of that
 in still, with $0.01$ in stopping, $0.6$ and $0.4$ for regular driving and cruise mode.
The differences of variance for gyroscope sensors are similar instead.
%, for stopping, moving,
% and the cruise mode are $0.0228$, $28.2620$, and $5.638$ respectively.
\emph{SmartLoc} continuously collect the sensory data from both accelerometer
 and gyroscope, if the vibration lies below the threshold, we consider the vehicle is stopped.
In our experiment, we find that \emph{SmartLoc} can differentiate moving and stopping activities
 precisely.

\section{Calibration by Landmarks}
\label{sec:landmarks}
\begin{figure*}[!ht]
\centering
\subfigure[Traffic lights\label{fig:red}]{\includegraphics[width=1.2in]{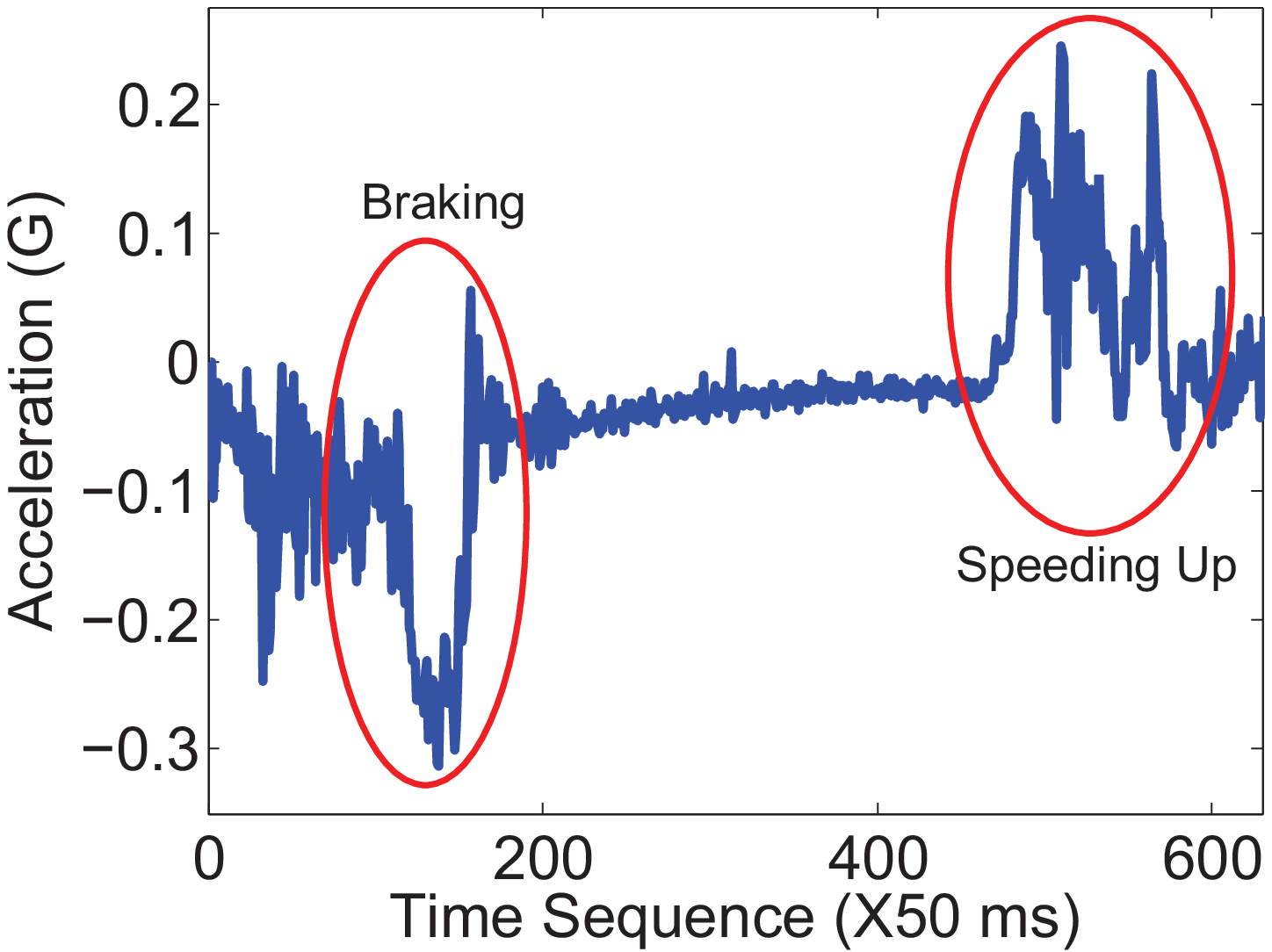}}
\subfigure[Centripetal force\label{fig:cen_for_turning}]{\includegraphics[width=1.3in]{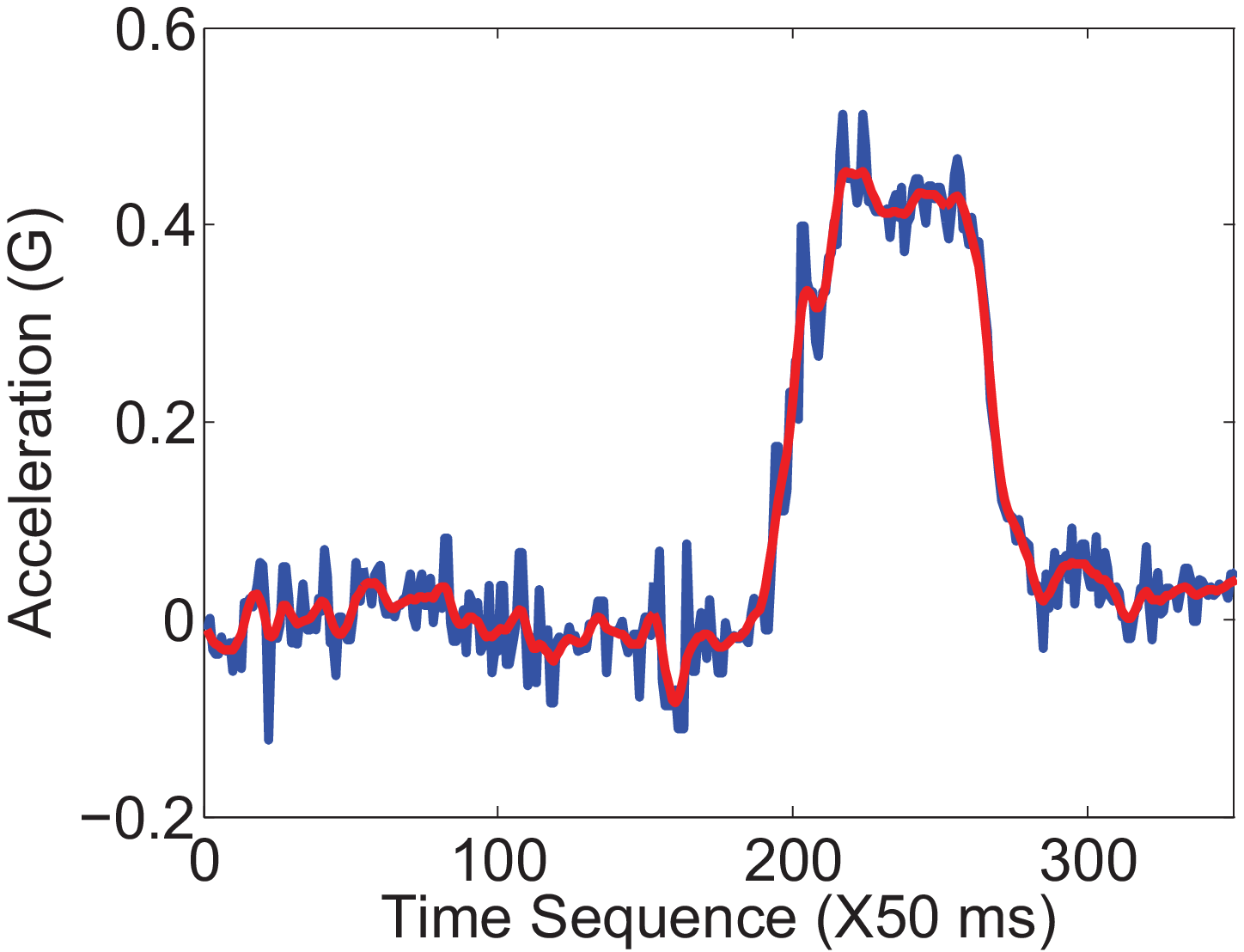}}
\subfigure[Angular velocity\label{fig:ang_v}]{\includegraphics[width=1.3in]{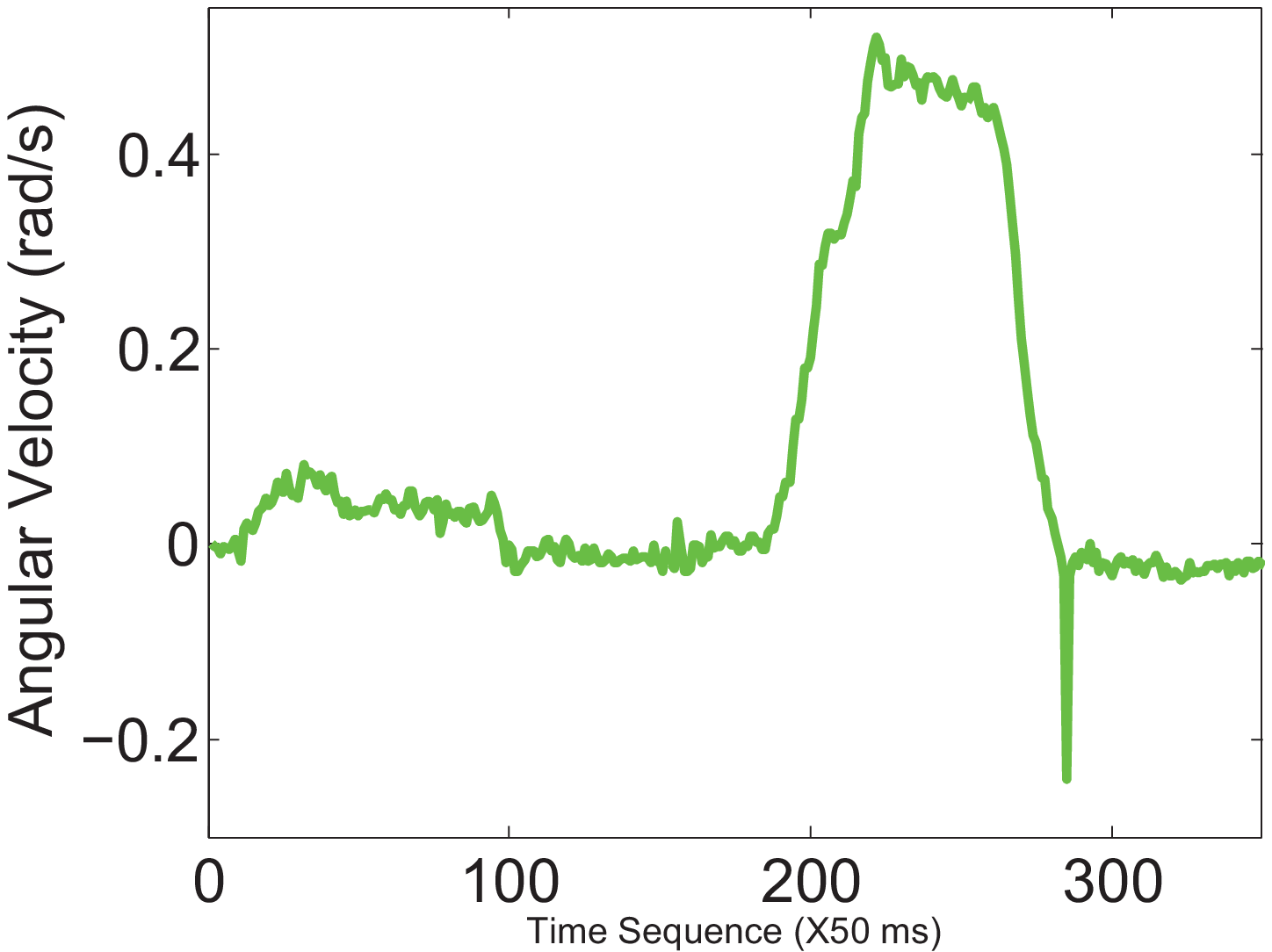}}
\subfigure[Magnetometer\label{fig:mag_turn}]{\includegraphics[width=1.3in]{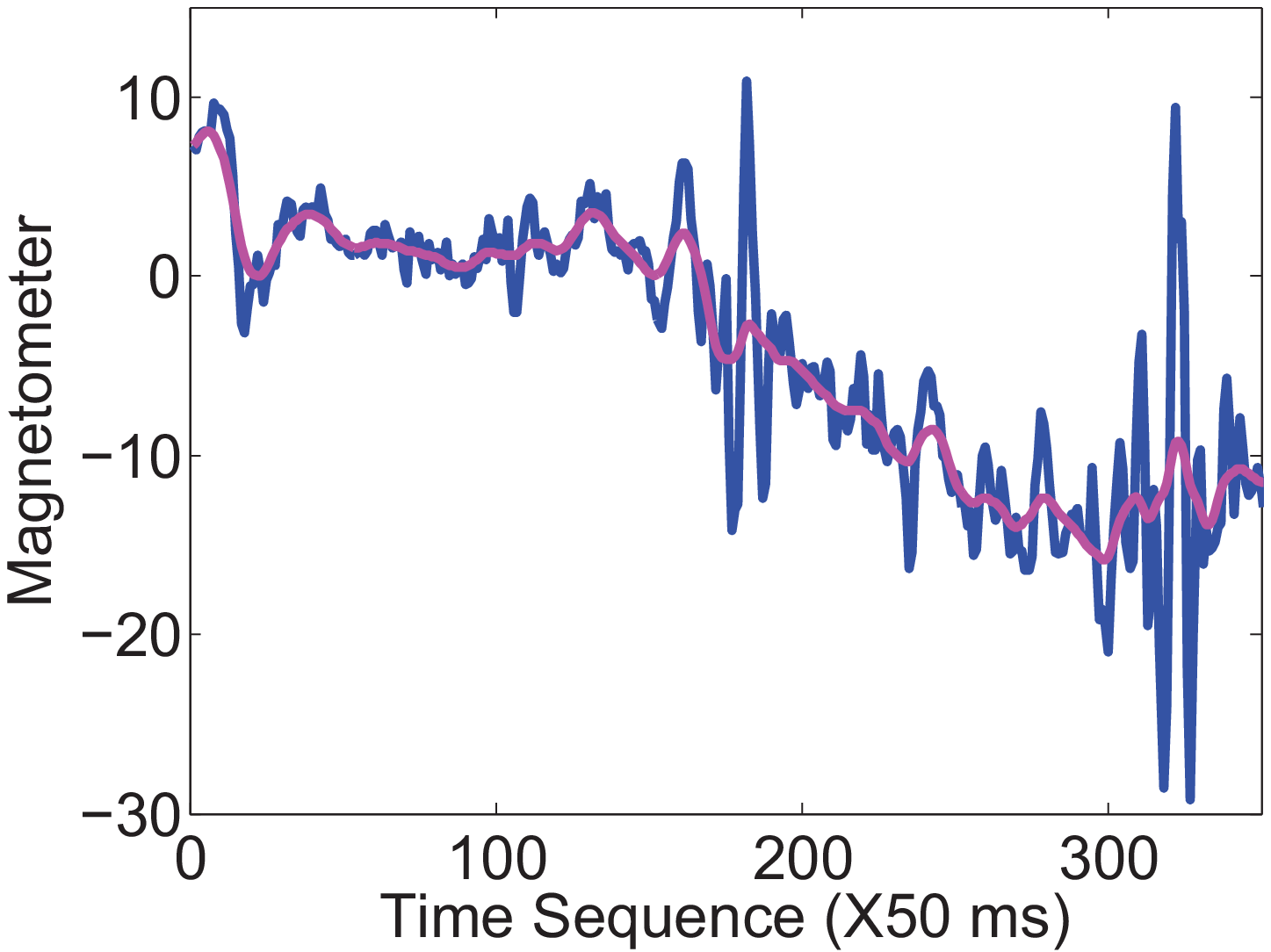}}
\subfigure[Bridge\label{fig:bridge}]{\includegraphics[width=1.3in]{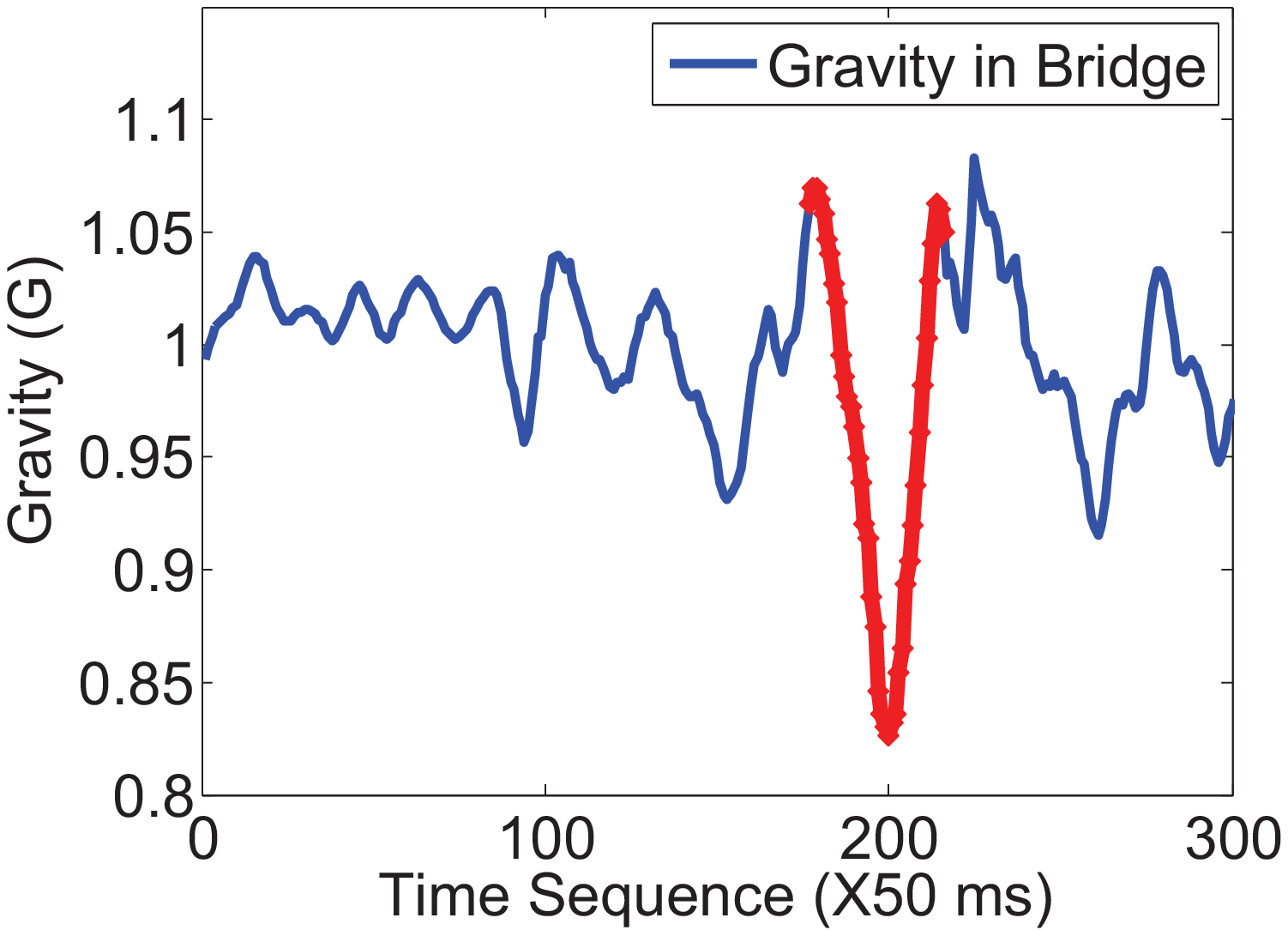}}
\vspace{-0.12in}
\caption{Pattern of the  sensor data collected in different
  road infrastructures when driving: (a) car stopping and crossing a
  traffic light; (b), (c), and (d) car turning $90^o$; and (e) car
  crossing a bridge.}\label{fig:pattern}
  \vspace{-0.11in}
\end{figure*}

As we have mentioned before, the road infrastructures, including
 tunnels, bridges, crossroads and traffic lights, cause large noises
 in the GPS data, which results in a large drift in the distance estimation if it is not treated
 rigorously.
In this work, we exploit the precise location of these
 infrastructures available in Google Map
 to calibrate the localization without any
 extra cost.\medskip

\textbf{Traffic Light:}
%Driving in many large well organized cities, such as Chicago, one often
%encounters lots of traffic lights.
When the vehicle stops due to
 the traffic lights and drives through crossroads,
 unique patterns appear in the readings of sensors (Figure~\ref{fig:red}).
Actually, when vehicles encounters traffic lights, the whole process can be
 divided into two phases, braking and speeding up respectively.
The acceleration falls below zero when the car brakes, reaching the lowest point
at the very moment when vehicle stops, and gets back to zero swiftly.
%As soon as the green light turns on, vehicle speeds up
%with the increasing acceleration.
However, in rush hours with terrible traffic, the location where cars stop may not be near the crossroad, but with a certain
distance from the crossroad. In this case, \emph{SmartLoc} adjusts the moving distance based on the estimated stopping location
from the empirical data, \ie, subtracting the distance from the car to
the crossroad.
However, since the distance between the car and the crossroad is determined
 by the traffic condition, it is difficult to measure the exact distance from the car to the crossroad.
The main approach adopted by \emph{SmartLoc} is to subtract the $\frac{n\cdot{L}}{2}$, where
 $L$ indicates the average length of a vehicle, and $n$ represents the current
 possible number of vehicles waiting for the green light.
According to our observation, the number of vehicles \emph{n} waiting before the traffic
 lights is related to the different time period.
In rush hours, the number of vehicles waiting is much larger, so that we assume such
 number follows normal distribution $n\sim{\mathcal{N}(\mu_t, {\sigma_t}^2)}$.

\begin{figure*}[hptb]
\centering
\subfigure[Turning\label{fig:turning}]{\includegraphics[scale = 0.25]{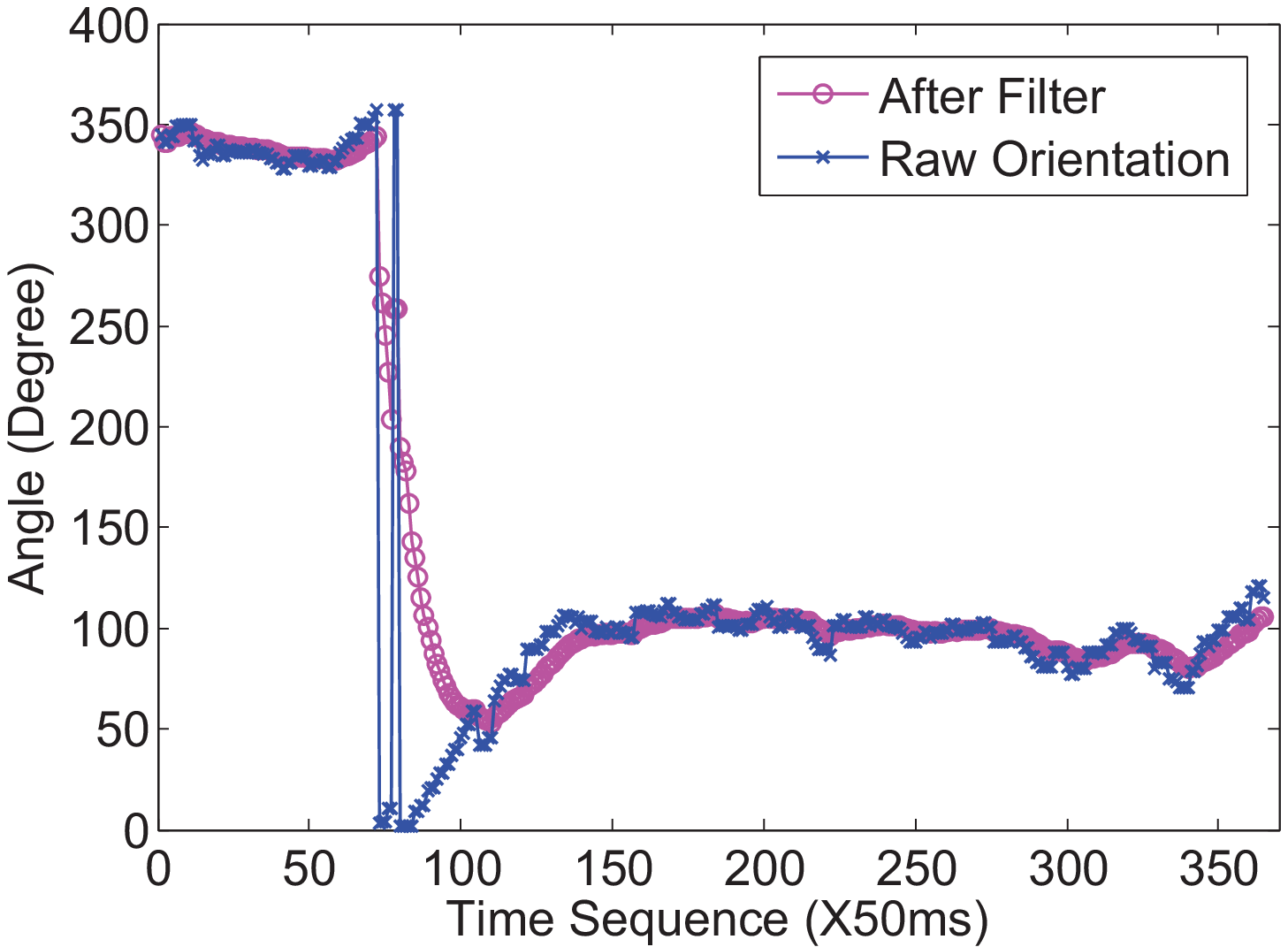}}
\subfigure[Changing Lane\label{fig:lane}]{\includegraphics[scale = 0.25]{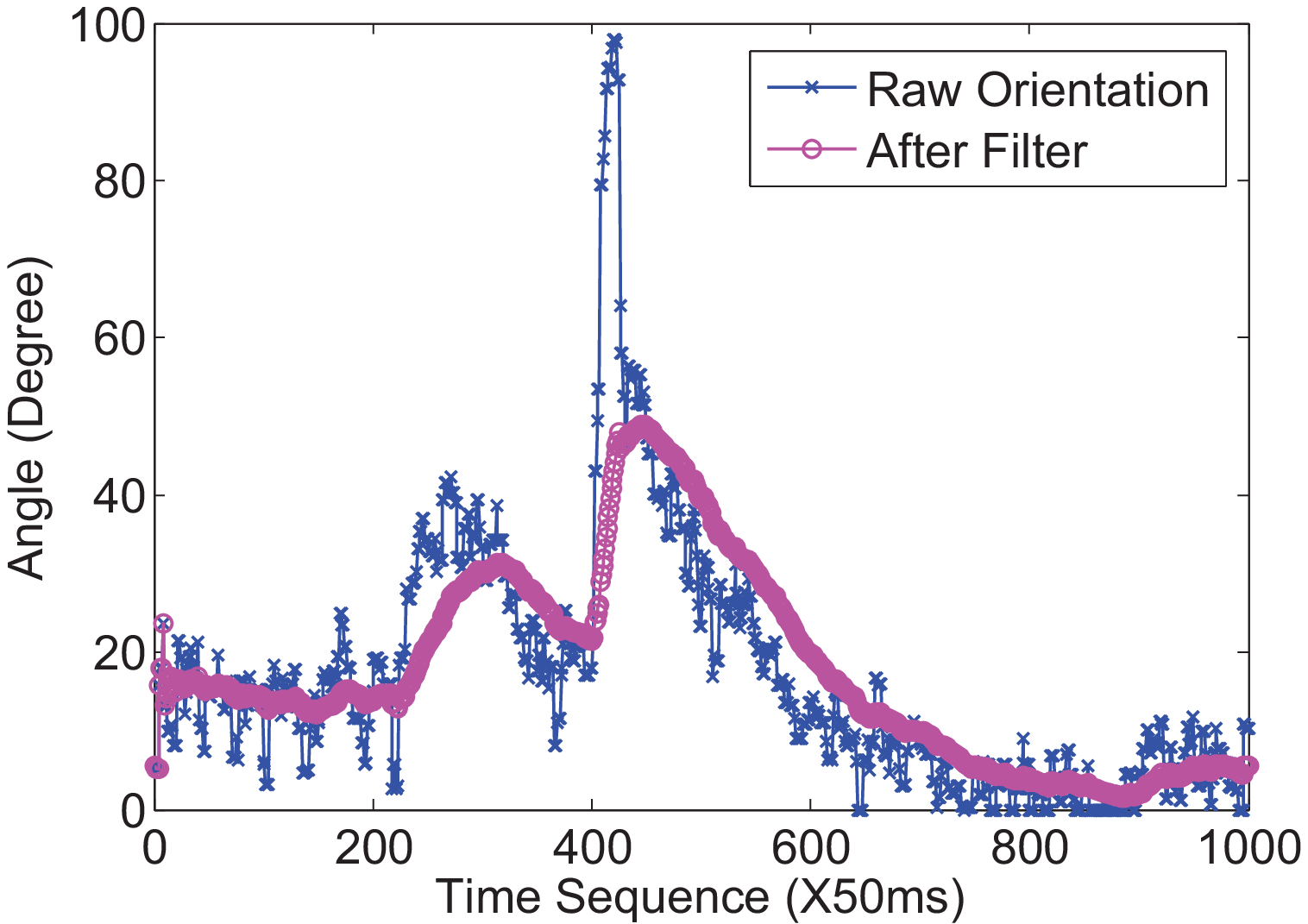}}
\subfigure[Driving Trace\label{fig:trace1}]{\includegraphics[scale = 0.25]{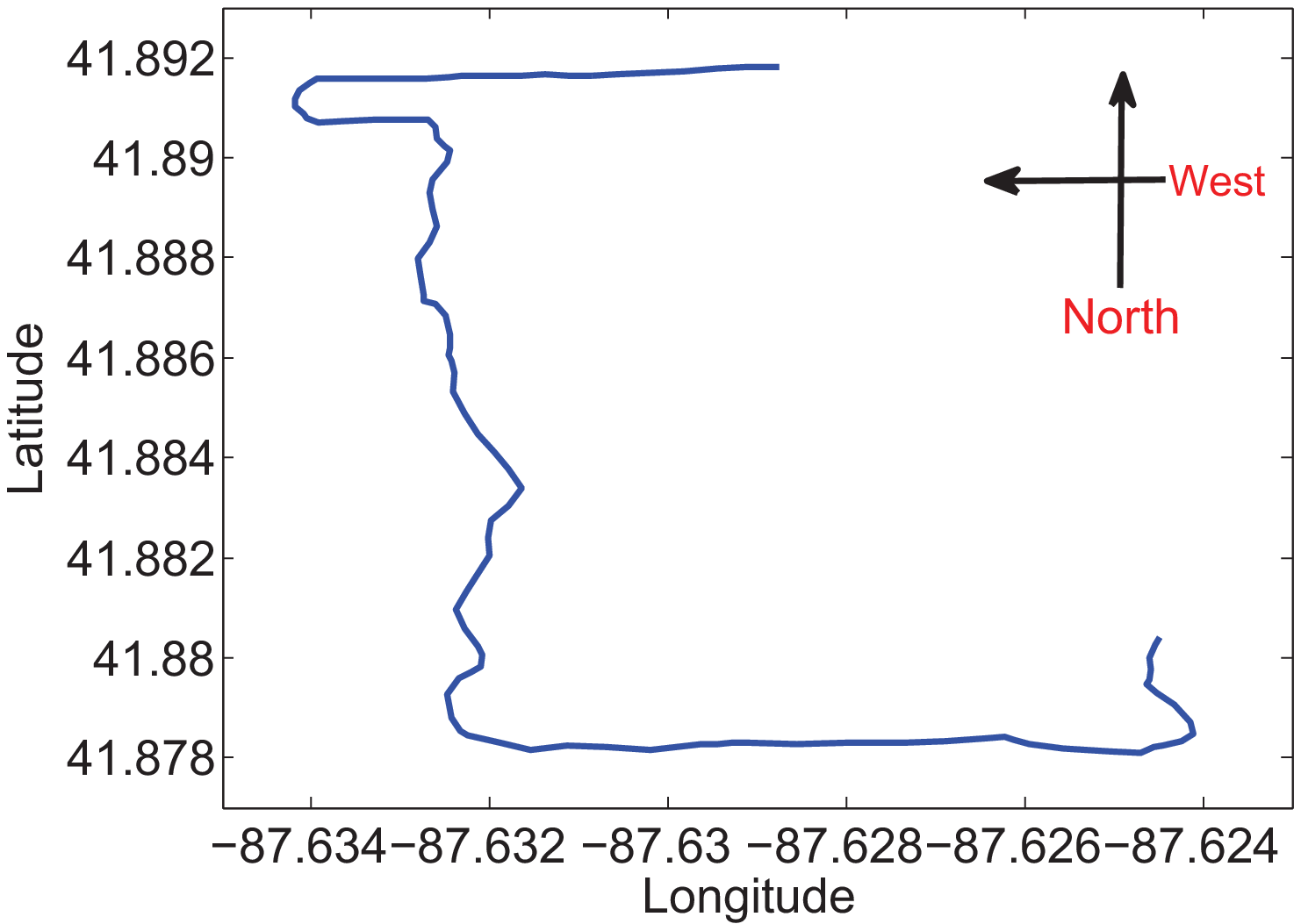}}
\subfigure[Estimated Orientation\label{fig:angle}]{\includegraphics[scale = 0.25]{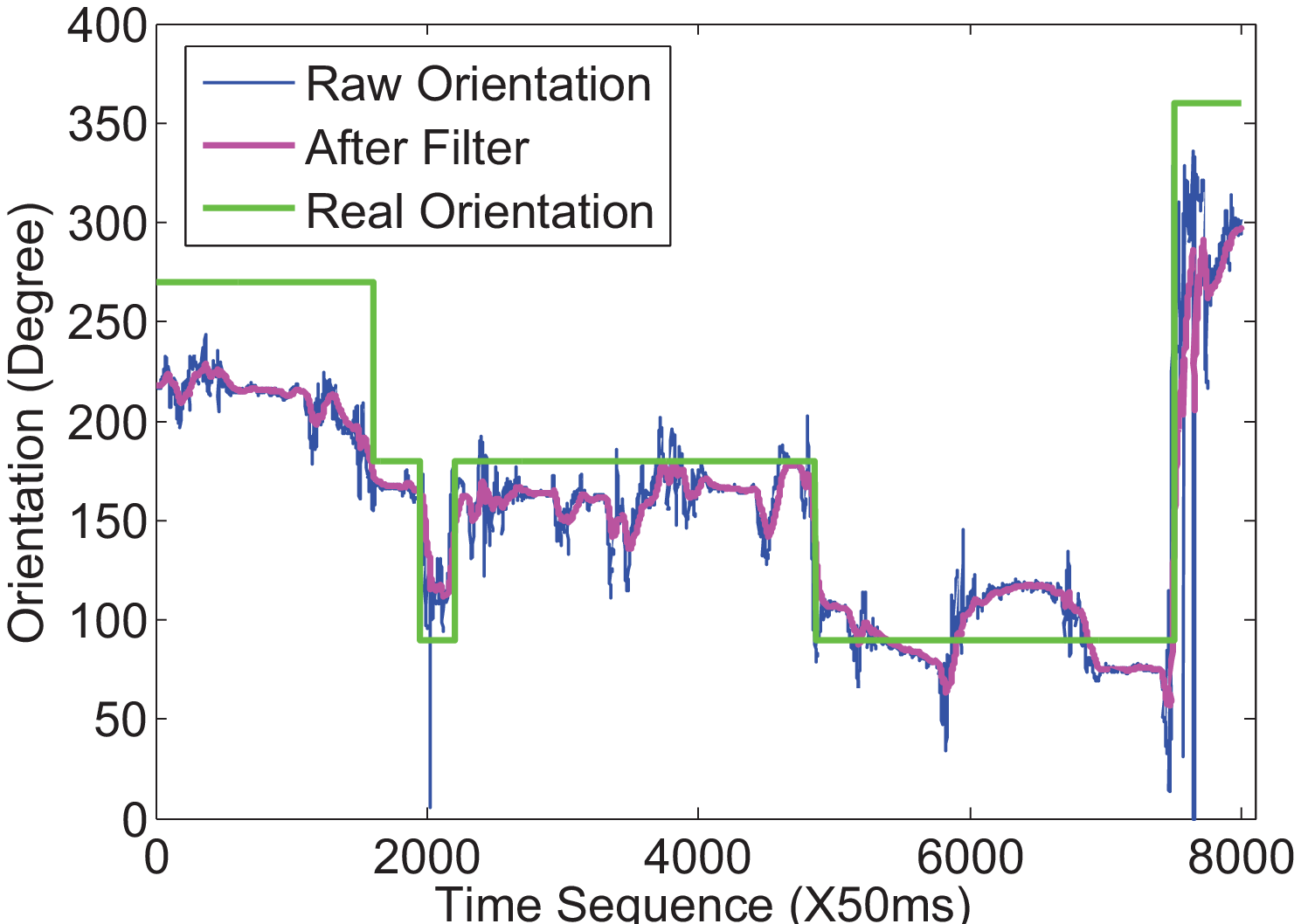}}
\vspace{-0.15in}
\caption{Turning or changing lanes, and Driving Trace\label{fig:turn_lane}}
\vspace{-0.15in}
\end{figure*}
\medskip
\textbf{Turning:}
Sometimes, vehicles may turn at intersections, which could be detected
 by sensors.
Figure~\ref{fig:cen_for_turning} indicates the
 centripetal force sensed by the accelerometer, and the scale of the
 acceleration depends on the speed at which the vehicle is
 turning.
Simultaneously, the angular velocity sensed by the gyroscope also
 reaches up to $0.5$ rad/s in our test case (Figure~\ref{fig:ang_v}), and
 the data from the magnetometer changes as well with a large
 fluctuation.
Finally, the orientation of the smartphone also changes
 approximately $90$ degrees when turning left or right.
Such angle change is observed along the axis in gravity direction,
 and the reading $0$, $90$, $180$, $270$ represent north,
 east, south, and west respectively.
Although the angle may not be accurate
 enough due to the large noise in the magnetometer
 (the maximum error we experienced was approximately $30^o$), we are still able to correctly determine the
 road segment to which the car is turning by calibration.
For example, Fig. \ref{fig:turning} shows a case when vehicle turns from
 the north, the angle is from about $350^o$  to $100^o$,
 which is east.
We also compare the measured angle differences for turning and lane changing (Figure~\ref{fig:lane})
 since lane changing can be wrongly detected as a turning.
In fact, the angle difference when a car changes its lane is much smaller than the one when a car make a turn.
In addition, we also calculated the standard deviation for
 the angle differences in lane changing, which is less than $10$.
Thus, distinguishing the turning and the lane changing is feasible.
Then, we conduct more studies on the driving orientation
 estimation.
Figure~\ref{fig:trace1} plots the raw trace of the vehicle
 achieved from the GPS with good signals, and Figure~\ref{fig:angle}
 illustrates the raw orientation generated only by the inertial sensors.
We employ moving average to cancel some noises and calculate the driving
 orientation, which matches the ground truth.

Other possible road infrastructures that a vehicle may experiences are
 bridges, and tunnels.
In our measurement, such patterns are more obvious and easier to be
 detected, mainly reflected in acceleration along the gravity direction,
 where the reading experiences a large and fluctuation when driving
 in a uphill or a downhill, as shown in Figure~\ref{fig:bridge}.

In fact, certain driving patterns, such as turing left or right and stopping for
 traffic lights or stop signs, can be more accurately detected and thus classified.
To classify other road infrastructures, we collect sensor readings of those patterns
 as the fingerprints, and then match the
 real-time sensor readings with the trained
 fingerprints.
To improve the classification and the matching accuracy, we rely on
 the coarse-grained estimation of the location from dead-reckoning first, and then we further use our predictive regression model (Section \ref{sec:regression}) to confine the search space: only the
 road infrastructures (stored fingerprints) $I$ within a certain distance $\delta$ from the
 estimated location $x$
 will be considered as the matching candidate for the real-time pattern $P$ achieved from the
 sensory data.
We select the infrastructure that maximizes the
\emph{weighted matching score}:
\begin{displaymath}
\alpha M(I, P)+ (1-\alpha) e^{-D(x,
  L(I))}
\end{displaymath}
where $M(I,P)$ is the matching score between the fingerprint
 of an infrastructure $I$ and the observed pattern $P$, $\alpha \in
 (0,1)$ is a constant, and $D(x, L(I))$ is the geodesic distance
 between the location $x$ and the location $L(I)$ of infrastructure $I$.
Then, the estimated location $x$ is updated as the location $L(I^*)$ of
 the infrastructure $I^*$ which maximizes the weighted matching score.

\section{Experiments and Evaluations}
\label{sec:experiment}
We conduct extensive evaluation of \emph{SmartLoc} in two different scenarios,
 both in downtown Chicago and suburb highways.
We test the performance in highways to evaluate the effectiveness and reliability
 of \emph{SmartLoc} replacing traditional GPS localization, so that it could save
 energy in navigation process.
In our evaluation, Samsung Galaxy S3 is mounted to
 the windshield, and we drive for over $100$ different
 road segments in downtown Chicago ranging from $1$km to $10$kms and
 over $30$kms in highway.
Since the inertial sensors provide the driving orientation, combined with
 driving distance from the location in last timeslot, the real-time location
 could be obtained.
Thus, the key problem becomes estimating the trajectory distance.
We evaluate the traveling distance,
 road infrastructure recognition, accuracy, and energy consumption.

\subsection{Trajectory Distance Estimation}
In trajectory distance estimation, we denote the trajectory distance
 in a timeslot as a \emph{traveling segment}.
Since the typical frequency for reliable GPS update in a smartphone (0.5Hz) is
 much lower than that of the sensors (1Hz-20Hz), we take duration for reliable
 GPS updating period as s timeslot.
We focus on the evaluation of
the trajectory distance estimation in two aspects:
(1) the accuracy in distance estimating in \emph{traveling segment};
and (2) the accuracy in final distance estimation of longer road segments.
Then, we analyze the performance in details in the rest of the section.

\subsubsection{Prediction in City Without Using Landmarks}
We first test \emph{SmartLoc} in downtown Chicago for over $30$ different roads,
 where some roads have reliable GPS signals and some not.
We separate these roads into more than $100$ road segments, whose sizes are determined
 by our evaluations presented in the rest of the section.
Before we describe the performance of \emph{SmartLoc} in metropolises, we have to admit
 that the GPS signals in downtown Chicago are relatively poor and time dependent.
Therefore, it is difficult to obtain the ground truth of all locations using smartphones,
 even if we adopt WiFi or GSM, fine grained location information are also
 hard to get.
In this case, we adopt the experiment in some areas with accuracy locations getting from
 GPS, and we remove some of the GPS information in these areas to simulate the missing
 good signal.
And we apply \emph{SmartLoc} to calculate the location in those removed road segments to compare with
 the ground truth.
Similarly, we analyze the performance of \emph{SmartLoc} in two phases as aforementioned.

\begin{figure}
%  \begin{minipage}[t]{0.66\linewidth}
    \centering
        \subfigure[Mean  error in each slot\label{fig:machine_learning_seg}]
        {\includegraphics[scale=0.25]{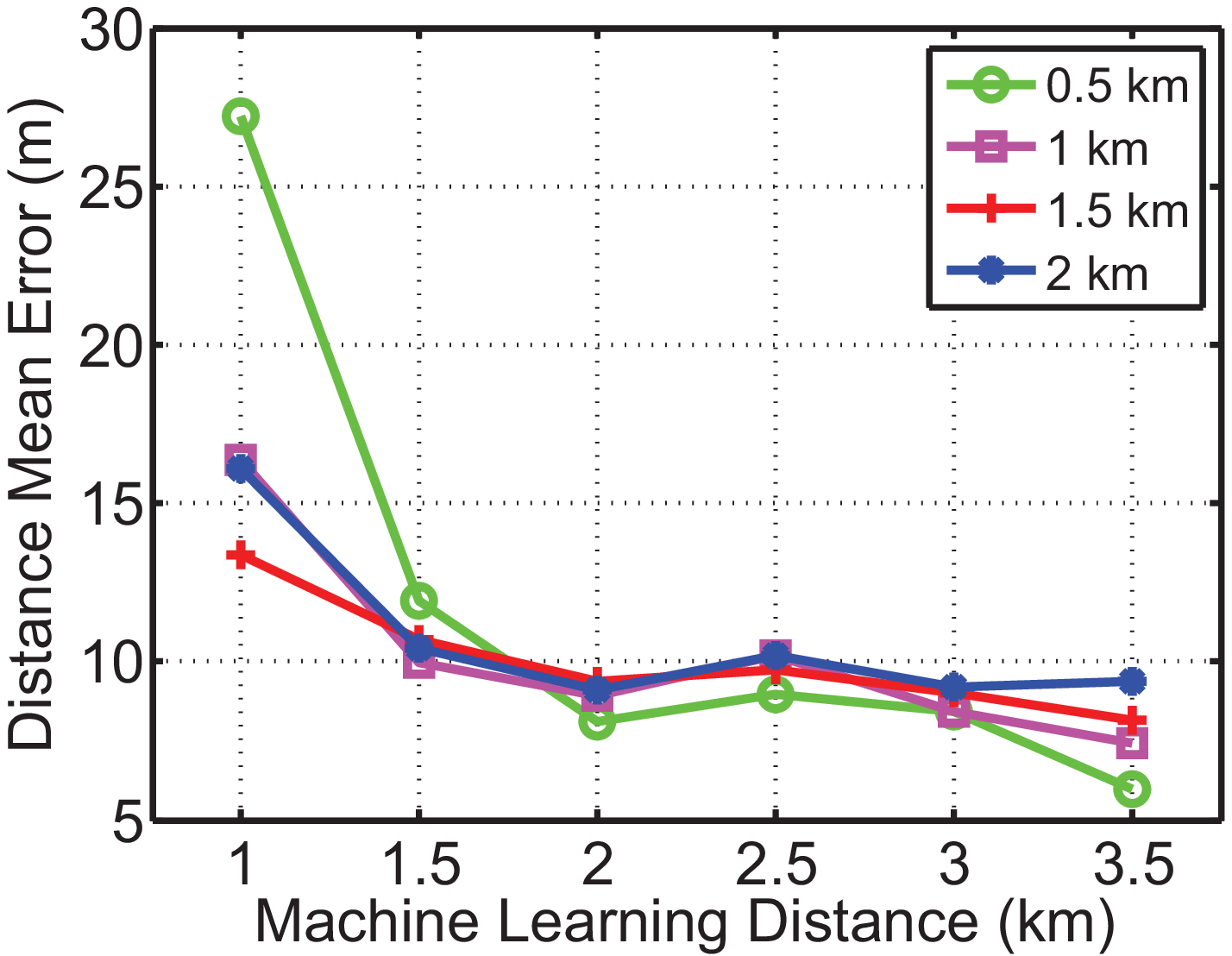}}
        \subfigure[Mean overall distance error\label{fig:machine_learning1}]{\includegraphics[scale=0.25]{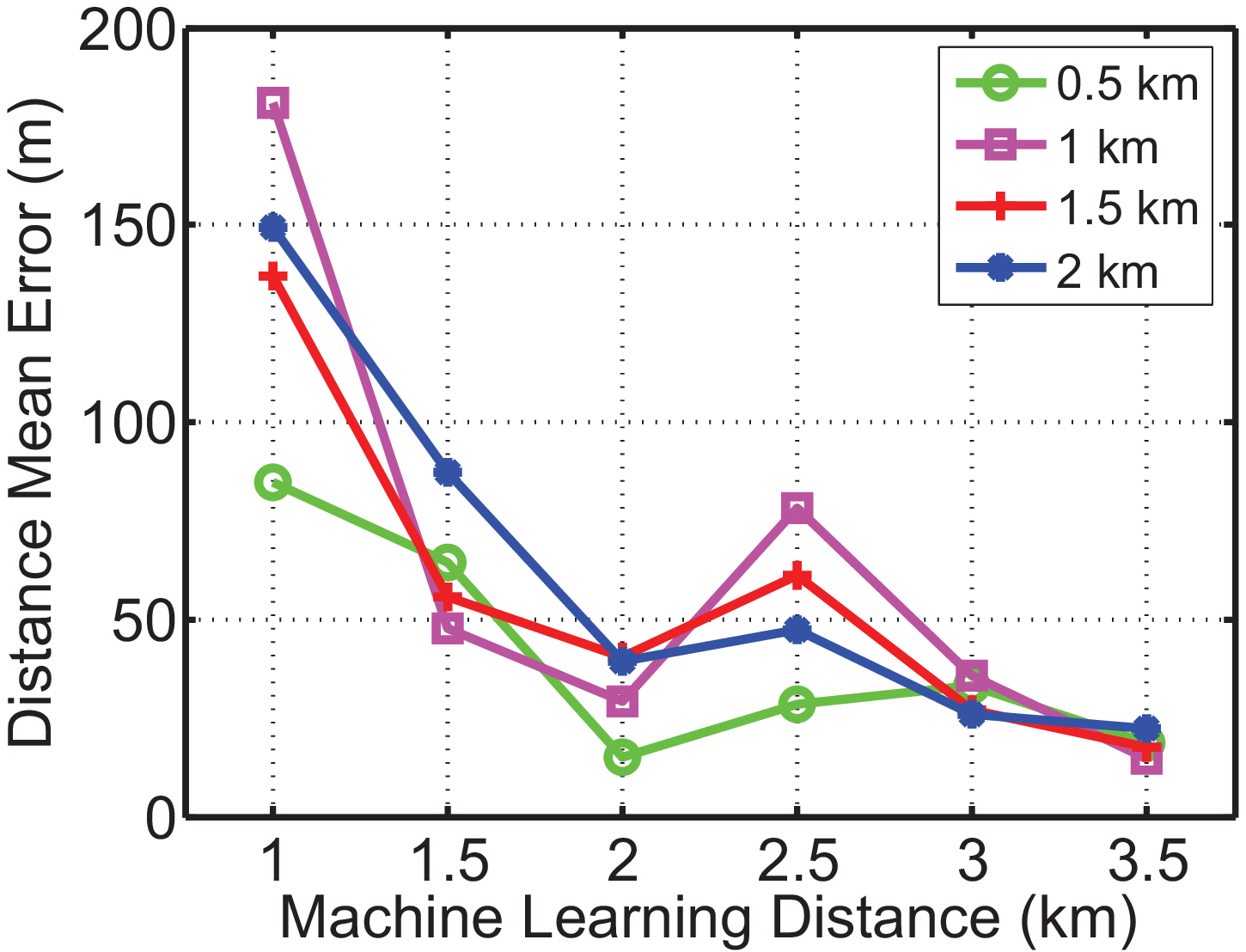}}
        \vspace{-0.15in}
        \caption{Accuracy vs. Learning Distance.}
        \vspace{-0.0in}
\label{fig:machine_local}
\vspace{-0.3in}
\end{figure}
Obviously, the driving habits and road conditions in a city are difficult to predict,
 and slight deceleration makes the predicted result deviate from the ground truth.
We first evaluate the reliability of \emph{SmartLoc} when different driving distances are used to train the system,
 ranging from $0.5$km to $3.5$km.
Generally speaking, the accuracy increases when the learning
distance increases as illustrated in Figure~\ref{fig:machine_learning_seg}.
In this figure, the X axis
 indicates the driving distance used for training our predictive regression model,
 and the Y axis represents the mean distance (between the actual location reported by
 the GPS and the  location estimated by our \emph{SmartLoc}) in every timeslot
 when we update GPS locations (\ie, every 2 seconds, or about every
 $22$m when driving at the speed $40$km/h).
This experiment measures the accuracy of the prediction when we drive for over
 four different road segments with length from $0.5$km to $2$km ($24$ different cases in total).
Due to the unstable driving activities, short road segments for training
 \emph{SmartLoc} leads to a large estimation error in each time slot.
When \emph{SmartLoc} learns only using the trace of $1$km, the mean error in
 every time slot in different scenarios is around $15$ meters, and
 the largest one is nearly $30$ meters.
When \emph{SmartLoc} trains our predictive regression model using a longer trace,
 the mean estimation error decreases  in all the test cases.
The smallest error is less than $6$m, which is less than half of the
 error when the training trace is  $1$km.
We also observe  that the error grows with the increase of the length
of the  test road segment in most scenarios.
For example, by  training \emph{SmartLoc} using a trace of $3.5$km,
  the mean error of the estimation in a $2$km road segment  is nearly
 twice of that when testing a $0.5$km road segment.

We then evaluate the  error on estimating the overall trajectory
 distance (Figure~\ref{fig:machine_learning1}) all the road segments and
 measure the error between the predicted distance and the ground truth
 distance for each segment (of all segments with
 distance from $0.5$km to $2$km) under different training traces.
If \emph{SmartLoc} learns the model for only 1km, the parameters in
 \equref{eq:dis_k_gps} cannot be computed accurately enough. Thus, the
  estimation errors increase to $180$m in all our
 tests.
When \emph{SmartLoc} learns enough samples, the parameters are much
 more reliable, and the  average accumulated error is far below
 $30$m, which is significantly better than the GPS in Chicago downtown.

\subsubsection{Prediction in City  Using Landmarks}
\begin{figure*}
%  \begin{minipage}[t]{0.66\linewidth}
    \centering
        \subfigure[Sensors Only\label{fig:MotionSensor}]
        {\includegraphics[scale=0.21]{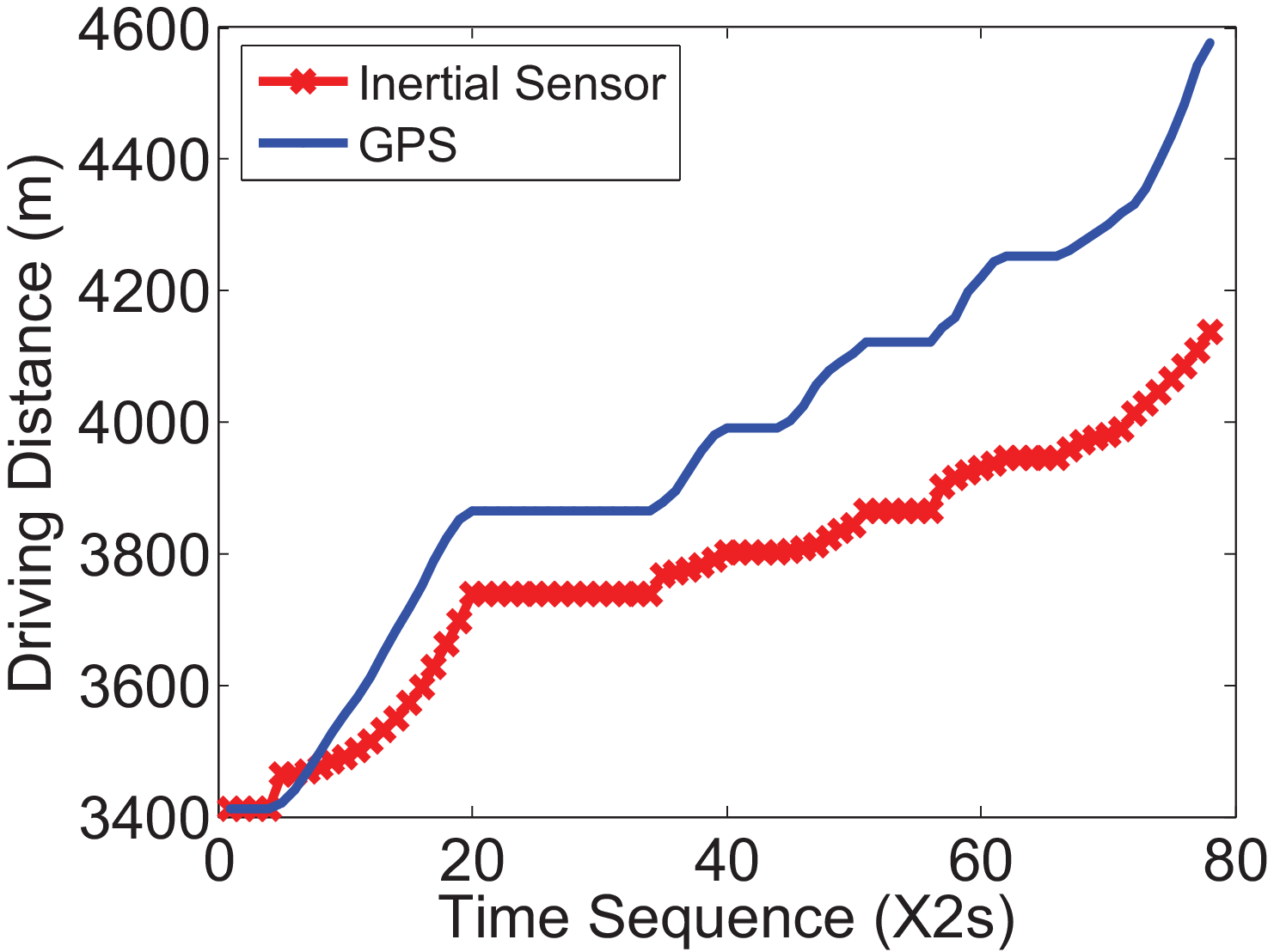}}
        \subfigure[Sensors and Traffic Lights \label{fig:Traffic}]
        {\includegraphics[scale=0.21]{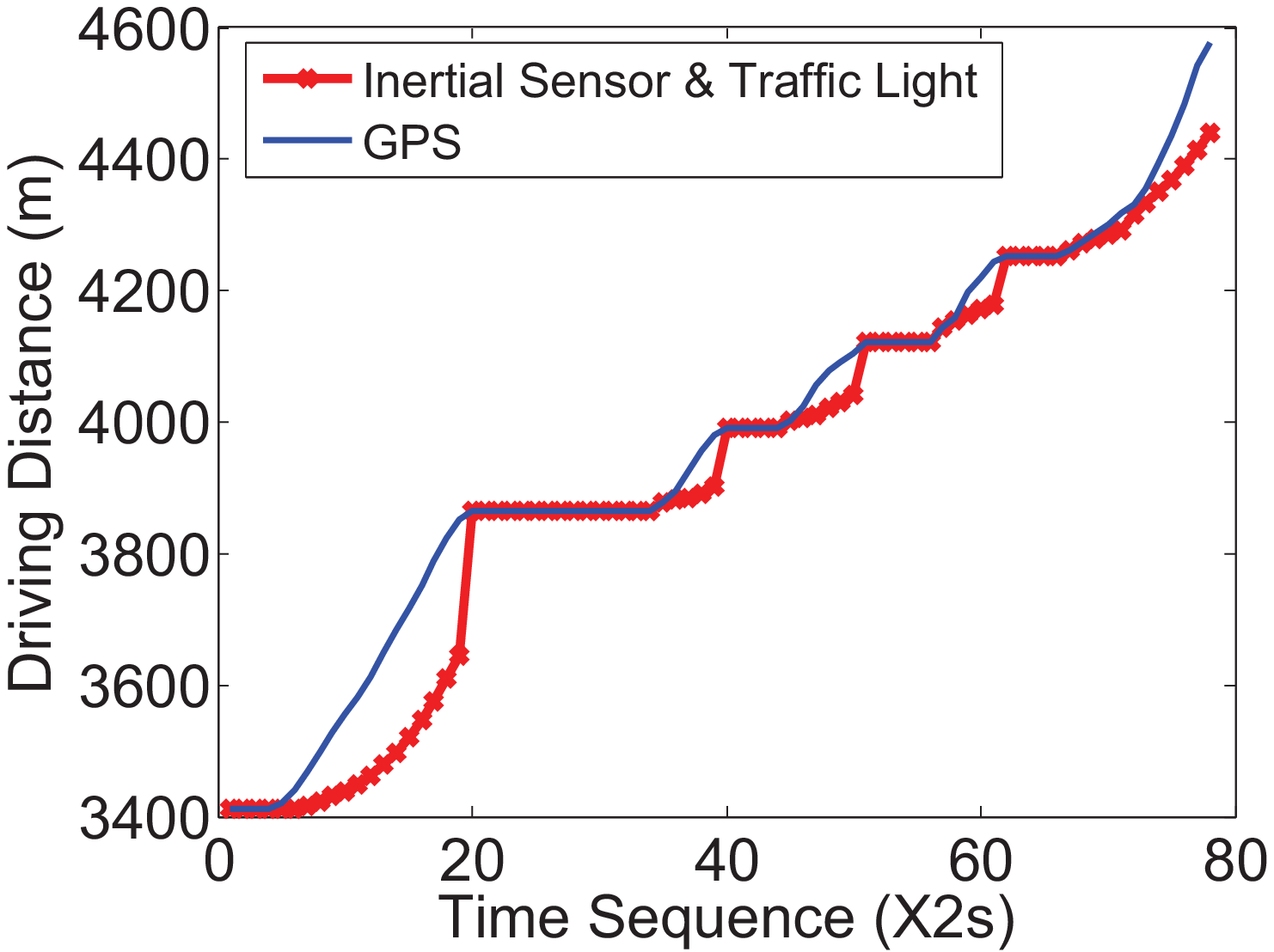}}
        \subfigure[Landmark vs No-Landmark\label{fig:landmark_comparison}]
        {\includegraphics[scale=0.21]{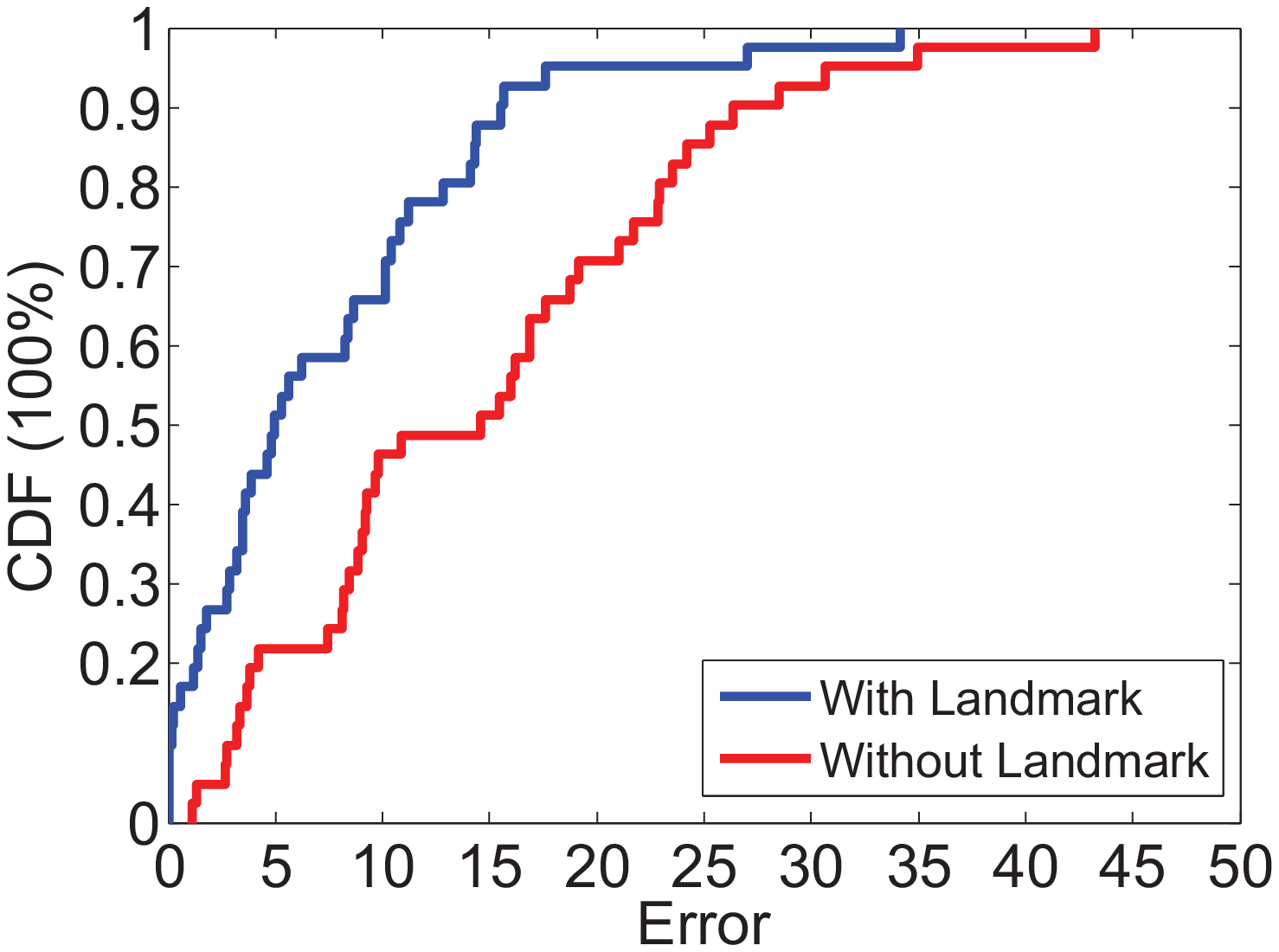}}
        \subfigure[SmartLoc\label{fig:smartLoc}]
        {\includegraphics[scale=0.21]{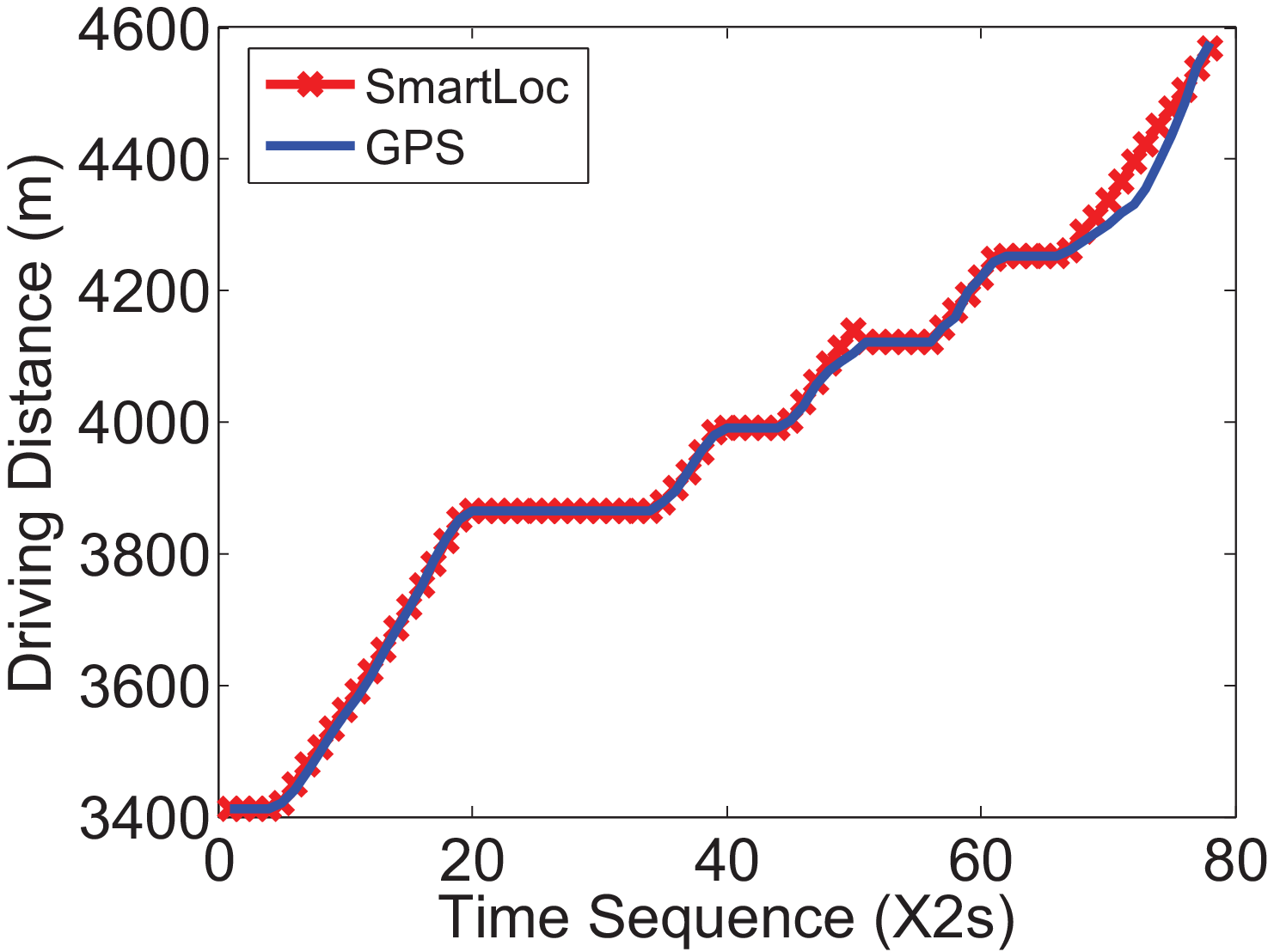}}
        \subfigure[Overall Comparison\label{fig:loc_gps_est}]
        {\includegraphics[scale=0.21]{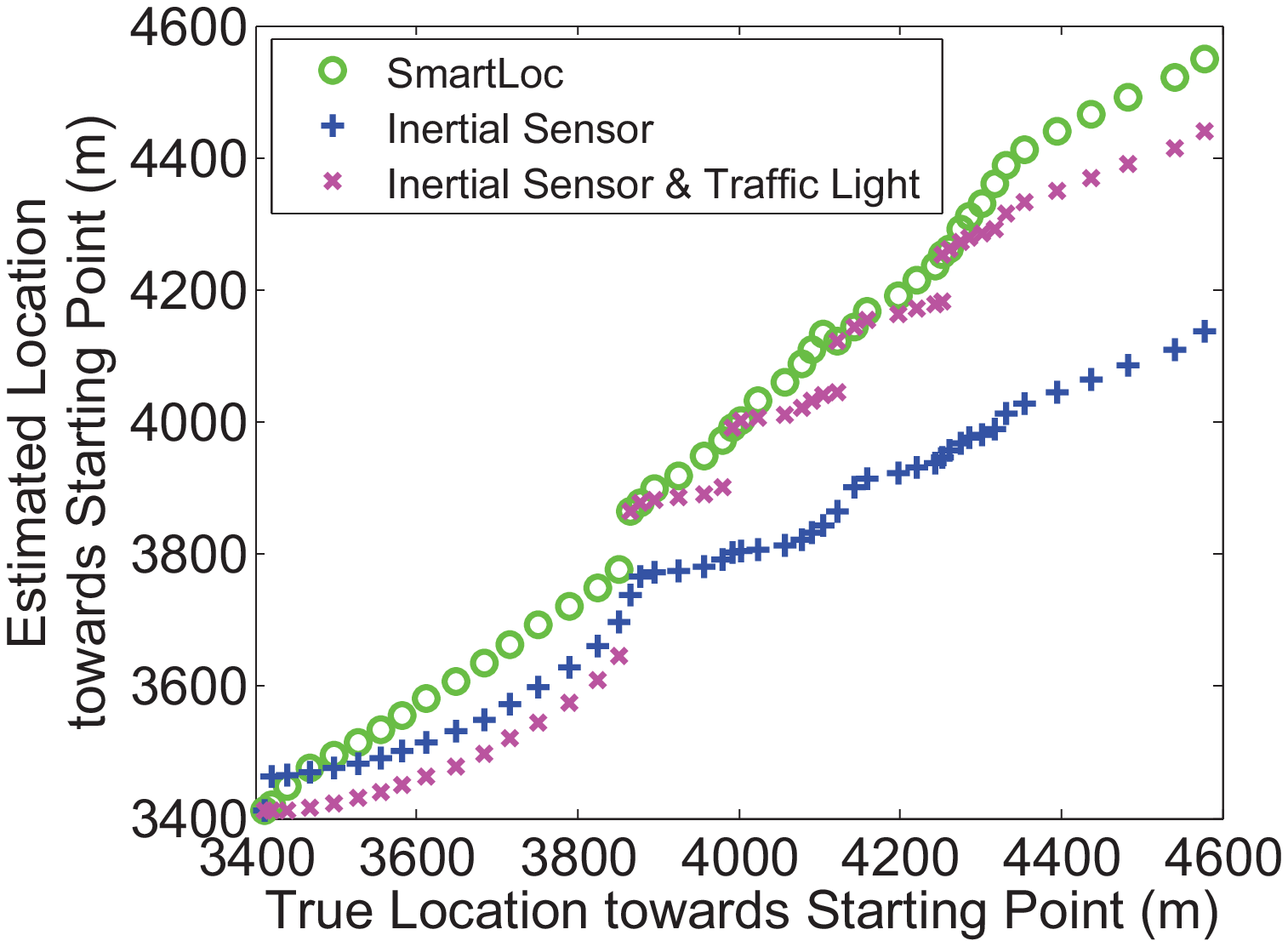}}
        \vspace{-0.1in}
        \caption{Distance prediction comparison among three methods and ground truth.}
        \vspace{-0.1in}
\label{fig:local_compare1}
\vspace{-0.15in}
\end{figure*}
\begin{figure*}
%  \begin{minipage}[t]{0.66\linewidth}
    \centering
        \subfigure[Error of driving distance in every time slot for three methods\label{fig:dif}]
        {\includegraphics[scale=0.33]{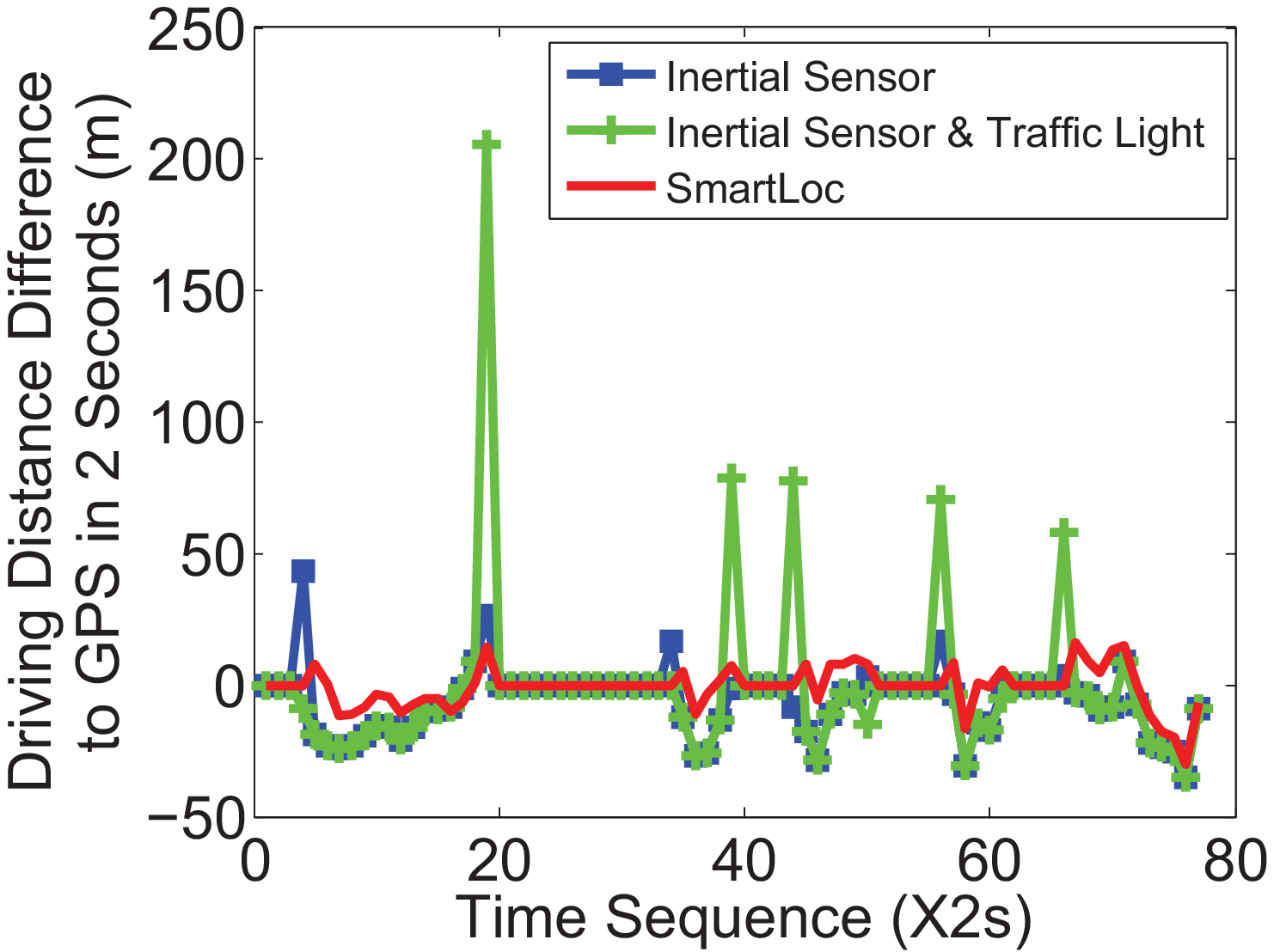}}
        \subfigure[Error of driving distance in every time slot for three methods\label{fig:dis_cdf2_seg}]
        {\includegraphics[scale=0.33]{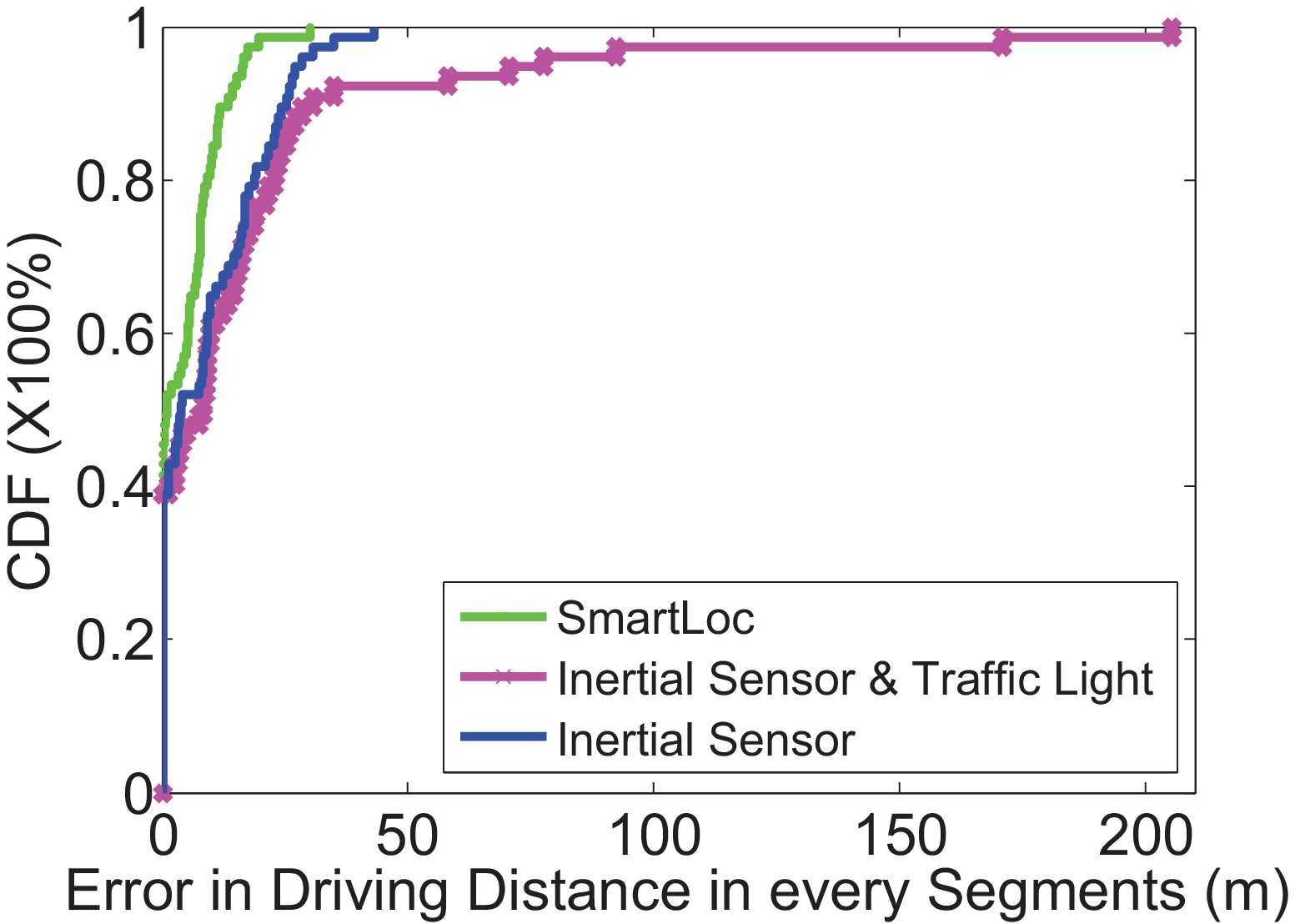}}
        \subfigure[Error of long distance estimation for three methods\label{fig:dis_cdf1}]
        {\includegraphics[scale=0.33]{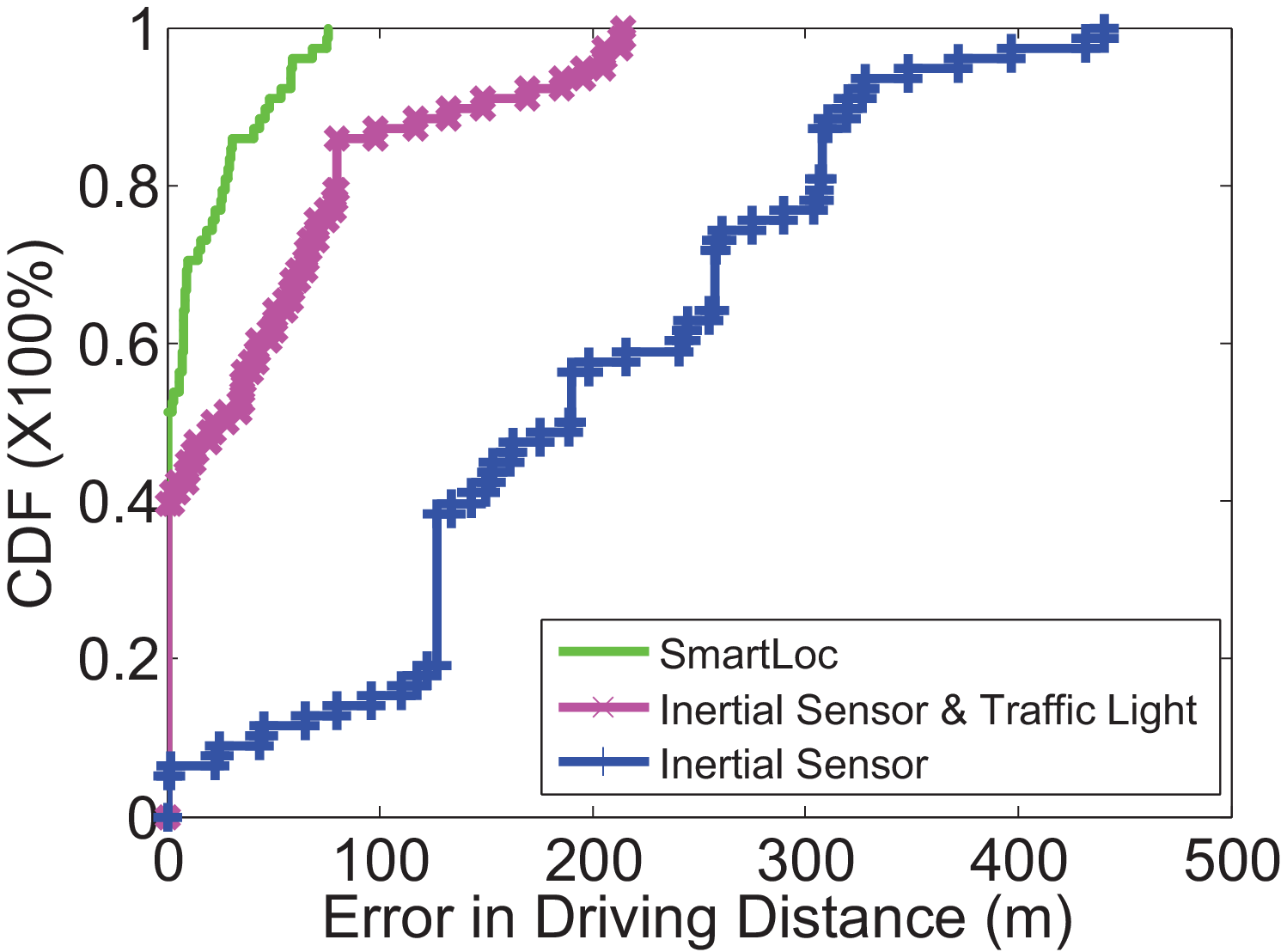}}
        \vspace{-0.1in}
        \caption{Comparison of three methods.}
        \vspace{-0.1in}
\label{fig:local_compare2}
\vspace{-0.15in}
\end{figure*}
\begin{figure}[h]
\centering
\vspace{-0.15in}
\includegraphics[scale=0.33]{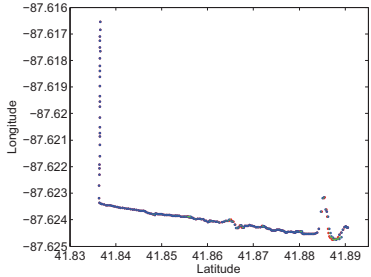}
\vspace{-0.15in}
\caption{Localization In the Street.}
\label{fig:michi_spot}
\vspace{-0.10in}
\end{figure}

\emph{SmartLoc} calibrates the location as soon as it
 detects specific patterns, especially traffic lights and
 turnings.
We test the performance of \emph{SmartLoc} in a real drive route with
 the calibration using landmarks, and the result is presented as Figure~\ref{fig:michi_spot},
which is a bird's-eye view of the driving trajectory.
The blue dots are the ground truth samples that we achieved from the GPS (where the GPS signals are good),
 the red dots are the predicted locations from our \emph{SmartLoc} with all calibration techniques,
 and the length of green lines denote the dimension of error.
In this figure, most of the red dots and blue dots are overlap with each other,
 which reflect the high accuracy in real downtown scenario.
%Both of them almost overlap with each other, and Figure~\ref{fig:landmark_comparison} also shows
% the significant improvement brought by our landmark calibration.

We then compare the performance of three different methods in detail: using
 inertial sensors only, using sensors and landmark calibration, and using \emph{SmartLoc} with all learning
 model and calibration.
In this experiment, we assume the first $3400$m is with
 reliable GPS signals, and the precise locations are
 accessible.
The estimation starts from $3400$m, and the first
 three figures in Figure~\ref{fig:local_compare1} indicate the driving
 distance from the starting point versus the elapsed time.

In Figure~\ref{fig:MotionSensor}, we conducted the experiment based on sensors only,
without any calibration or noise canceling. The double integration on acceleration leads to
the final deviation of over $400$m after driving about $1200$m.
When the road pattern detection is introduced, the location is
 calibrated when \emph{SmartLoc} senses the road infrastructure pattern. During the
 same experiment, our vehicle crossed $5$ traffic lights in total, and
 successfully detected all 5 traffic lights.
The estimated locations are all then adjusted accordingly.
The error in Figure~\ref{fig:Traffic}  is still high, especially in the
crossroads.
Surprisingly, after combining our predictive regression model,
\emph{SmartLoc}'s result almost coincides with the ground truth, as shown in Figure~\ref{fig:smartLoc}. For
the first $900$m, the curve of \emph{SmartLoc}  nearly overlaps
with the curve of the ground truth. For the first $450$m, the vehicle
passes three crossroads with all green lights, and the error is less than
$20$m in most of the time. After the final traffic lights, the
vehicle has to drive at a relatively low speed because of the road
construction.
The predicted distance consequently deviates from the
 ground truth a little, but at the end of the road, the errors remain
 small.
We plot all the estimated distances by three methods in
 Figure~\ref{fig:loc_gps_est}, with the X axis being the ground truth
 distance and Y axis being the predicted distance, \ie, the perfect prediction will have a diagonal line.
\emph{SmartLoc} results are distributed almost along the diagonal line, and pure
 sensor approach  deviates greatly.

The deviation of the results from the ground truth comes from the accumulated
errors from all time slots. Based on the previous experiments, we plot
the error in every time slot in Figure~\ref{fig:dif}.
\emph{SmartLoc} with landmarks calibration
 has the smallest mean error of the estimated locations for all time slots:
 $90\%$ of them are lower than $20$m from the CDF in
 Figure~\ref{fig:dis_cdf2_seg}.
The other two approaches have larger errors, and the last figure
 describes the CDF of the total driving distance error.

\subsubsection{Prediction in Highway}
\begin{figure*}
%  \begin{minipage}[t]{0.66\linewidth}
    \centering
        \subfigure[Traveling Distance\label{fig:highway_dis}]
            {\includegraphics[scale=0.25]{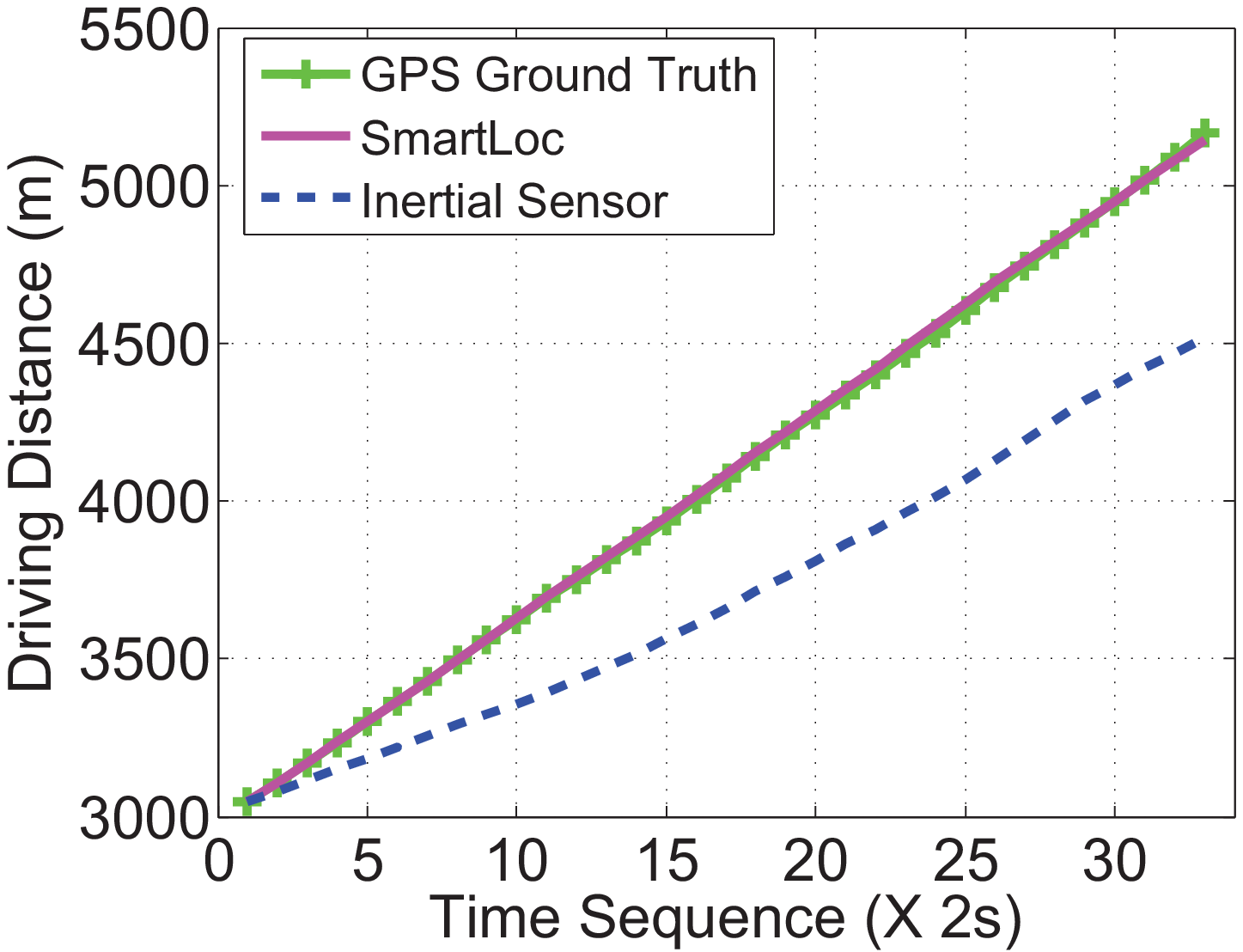}}
        \subfigure[Absolute error\label{fig:highway_seg}]
            {\includegraphics[scale=0.25]{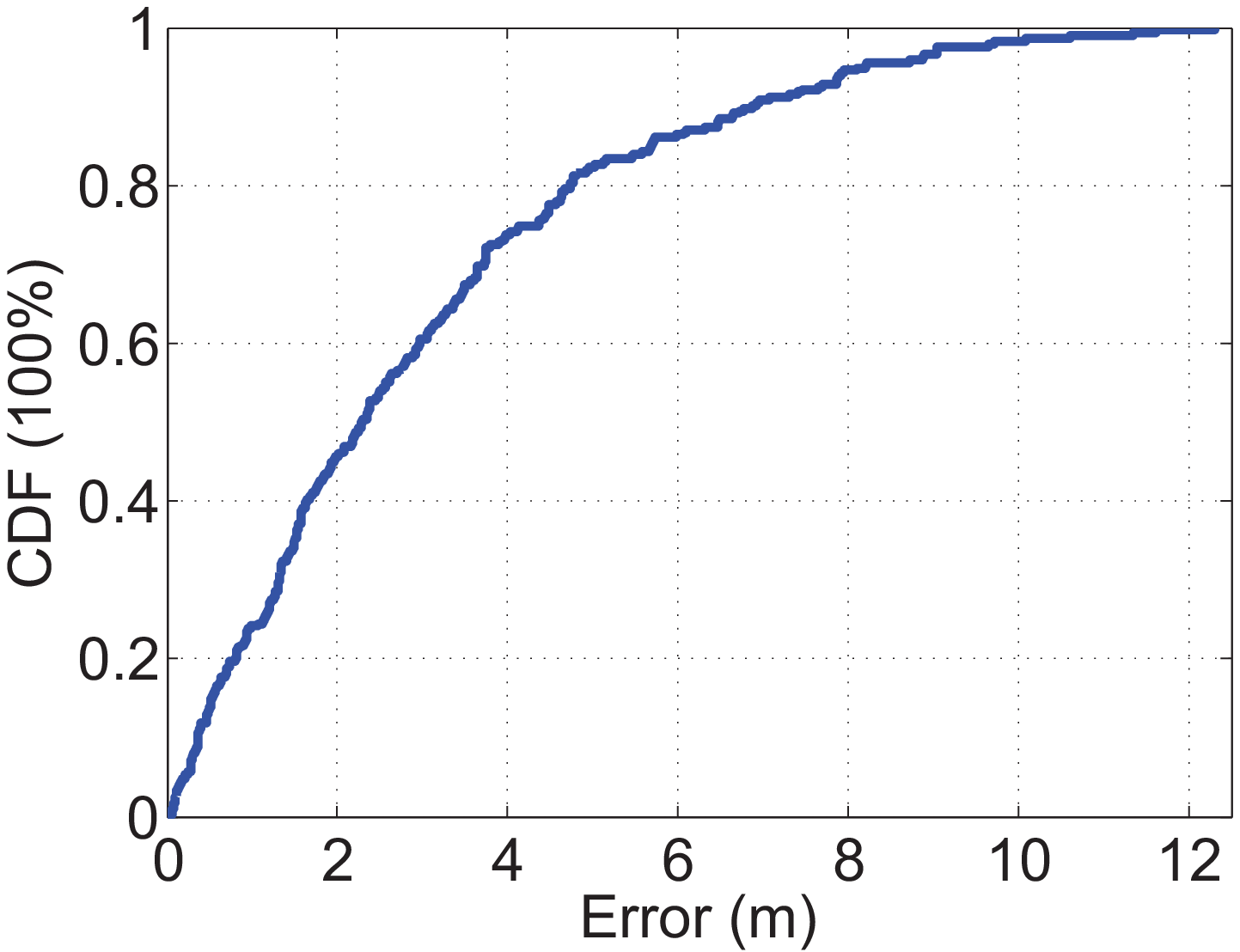}}
        \subfigure[Relative error\label{fig:highway_seg_cdf}]{\includegraphics[scale=0.25]{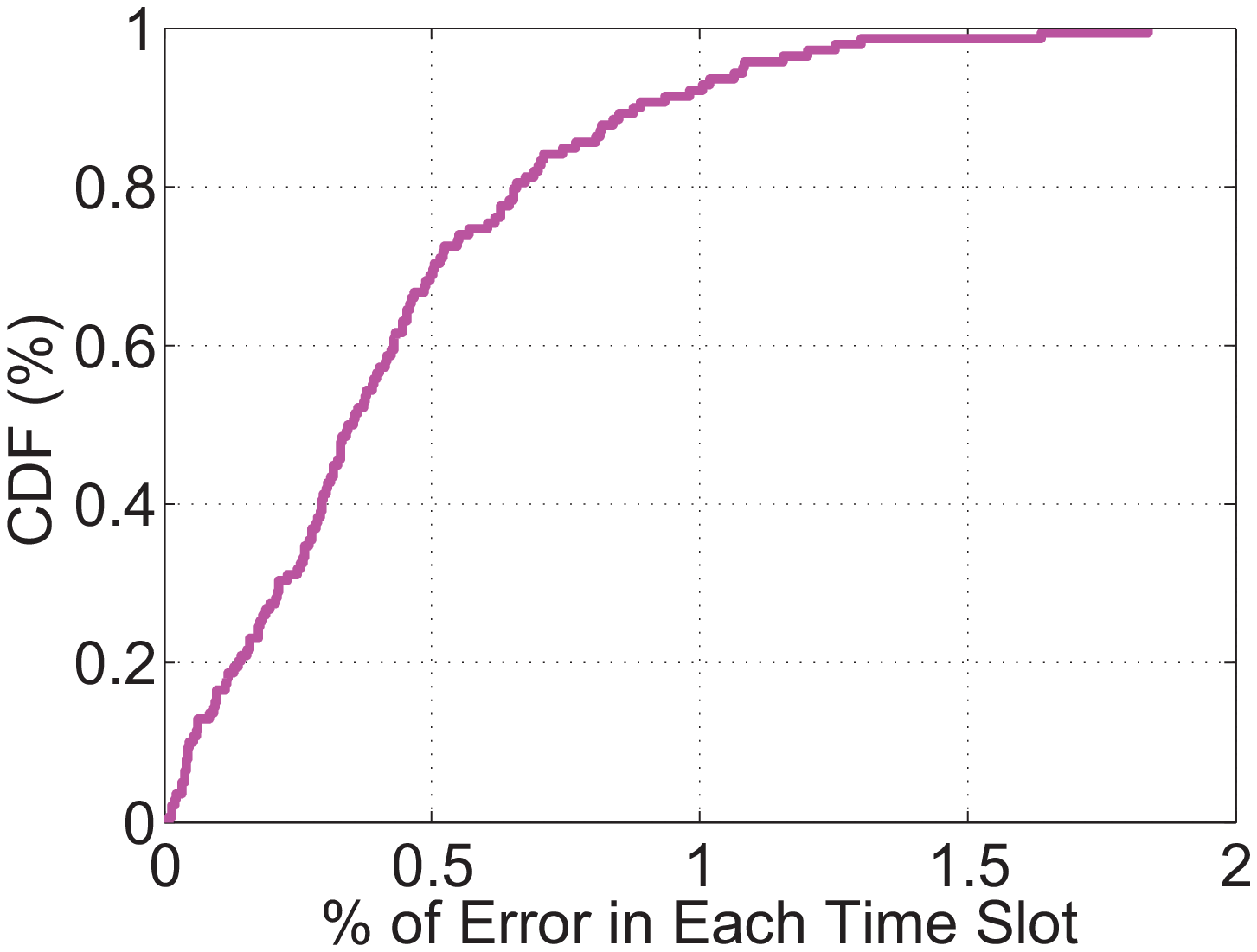}}
        \subfigure[Long Distance Estimation\label{fig:highway_dif_cdf}]{\includegraphics[scale=0.25]{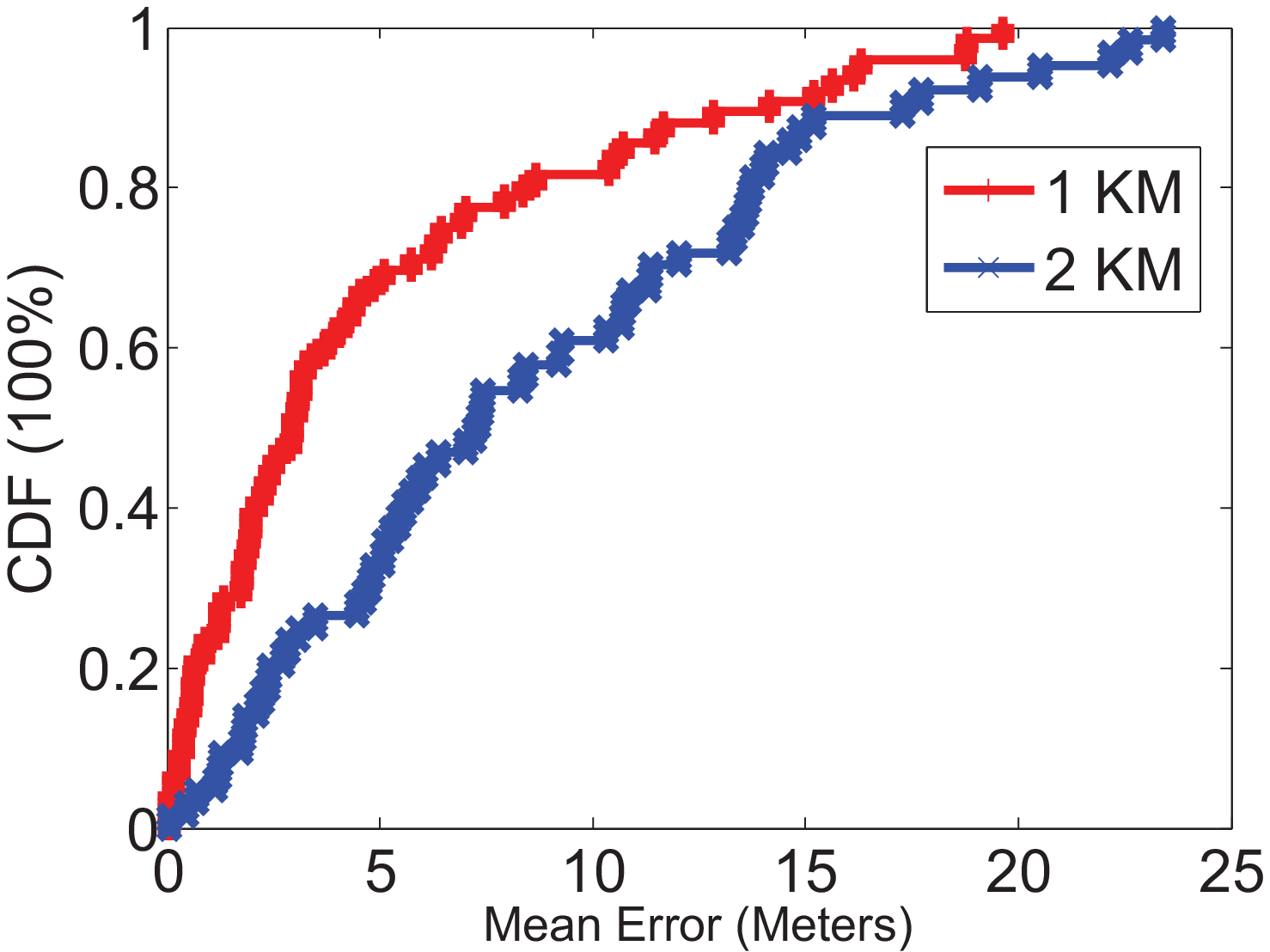}}
        \vspace{-0.1in}
        \caption{Traveling in a highway.}
        \vspace{-0.3in}
\label{fig:highway}
\end{figure*}
In addition, we test the performance of \emph{SmartLoc} on the highway to evaluate the
 probability of replacing traditional GPS to save energy.
In the highway, GPS signals are almost always good, so the GPS data served as
 the ground truth only in this evaluation.

We drive over $10$ different highway segments with total distance being
 over $60$km (with driving speed $100$km/h-$120$km/h approximately).
The smartphone has access to the precise
 location information from the GPS, which is updated every $2$ seconds.
Meanwhile, we collect the readings from the sensors and train our predictive regression model for $3$km.
Then, we predict the traveling distance for the next $2$km and compare the distances from the ground truth, \emph{SmartLoc}
 and the pure sensors.

Figure~\ref{fig:highway_dis} illustrates the comparisons of driving
distance estimation using \emph{SmartLoc} (with sensors) and the GPS.
The ground truth (GPS readings) is plotted by the green curve.
It is obvious that the error between pure sensor-based solution and the
 ground truth is becoming larger along the time, which is due to
 the accumulated errors without any calibration.
By using our predictive regression model, \emph{SmartLoc} calculate
 suitable parameters and apply them into the prediction. The
 estimation errors gets much smaller after then.
Figure~\ref{fig:highway_seg} indicates that the largest error
 is only $12$m among the $10$ different highway segments (each
 of length $2$km), and in over $80\%$ cases, the errors are less than $5$m.
Compared with the actual distance extracted from the ground truth (Figure~\ref{fig:highway_seg_cdf}), at over $95\%$ locations (among all
 locations where GPS location can be extracted), the errors are less than
 $1\%$ of the actual driving distance, and the largest error is less
 than $2\%$ of the actual driving distance.
We also notice that the accuracy
of the prediction decreases with the
increase of the driving distance.
We predict the driving distance for both $1$km and $2$km after
 taking the data of the first $3$km to build the model.
In our experiments, $80\%$ of the prediction error for both (1km) and (2km) cases
 are less than $10$m and $15$m respectively, and even the largest error fall within
 $19.8m$ and $23m$ as plotted in Figure~\ref{fig:highway_dif_cdf}.

However, based on the evaluation, we discover that the estimation
 results cannot maintain high accuracies for a long distance even in highway.
The main reason comes from the user dependent driving behaviors and the
 unpredictable special conditions, such as traffic jam.
We also consider that \emph{SmartLoc} has a better estimation accuracy when the
driving speeds remain stable, and when the  driving speed fluctuates frequently, the
 error of \emph{SmartLoc}'s predicted results still in an acceptable range.
Calibrating the location periodically is a feasible way to improve the
 location accuracy in real life applications, which is also an alternative
 to replace traditional GPS to save energy in the highway.

\subsubsection{Evaluations Analysis}
Based on the evaluation results presented in this section, an
 obvious conclusion is that \emph{SmartLoc} provides precise driving
 distance estimation in certain scenarios.
In every time slot, the driving
 distance is estimated from the current sensor data as well as
 our predictive regression model.
Suppose the error (denoted as $D_i$) in the estimation
 of each time slot $i$ follows normal distribution: $D_i\sim{\mathcal{N}(\mu,
  \sigma^2)}$, with mean $\mu$ and  variance $\sigma^2$.
Then, the estimation of the total traveling distance $S_t$ in $t$
 timeslots is the summation of the traveling distance in all time slots:
$S_t = {\displaystyle\sum\limits_{i=1}^{t} D_i}$.
In this case, the error, from a long term perspective, will be
 accumulated.
Obviously, $S_t \sim{\mathcal{N}(t\mu,  t\sigma^2)}$.
The variance of the variable $S_t$ will be $ t\sigma^2$.
Thus, the mean error increases along the time, which
 leads to the conclusion that it is difficult to predict the traveling distance
 precisely in a long term, although sometimes the deviation in some continuous
 timeslots may  be neutralized.
For a given error bound $\delta$,
$Pr(S_t \geq \delta)$ is higher when $t$ is larger.

\vspace{-0.1in}
\subsection{Localization in the City}
\begin{figure}
%  \begin{minipage}[t]{0.66\linewidth}
    \centering
        \subfigure[Error of location estimation in every sampling timeslot\label{fig:total_block_seg_cdf}]
        {\includegraphics[scale=0.25]{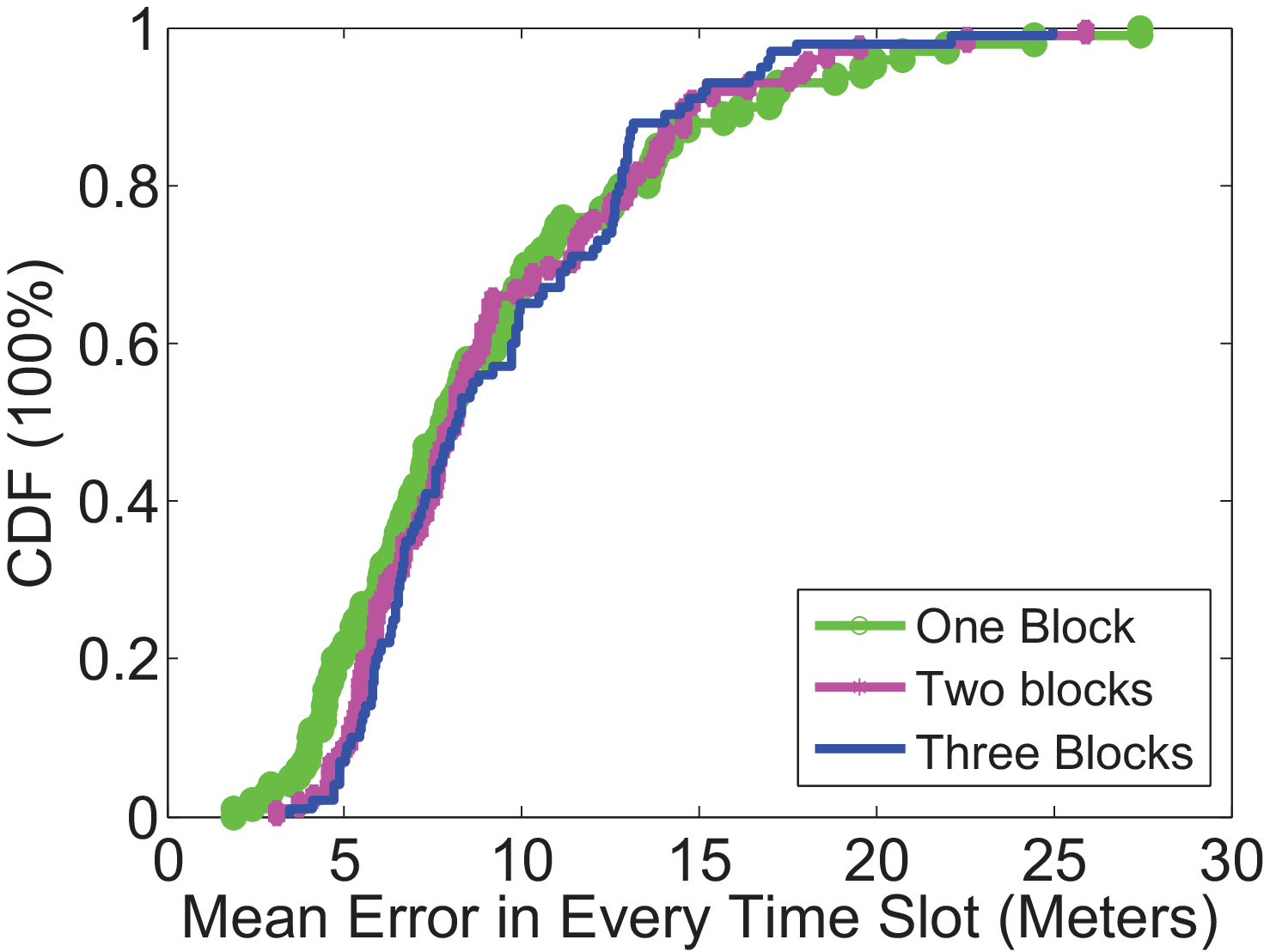}}
        \subfigure[Error in locating the final destination in different blocks\label{fig:total_block_cdf}]
        {\includegraphics[scale=0.25]{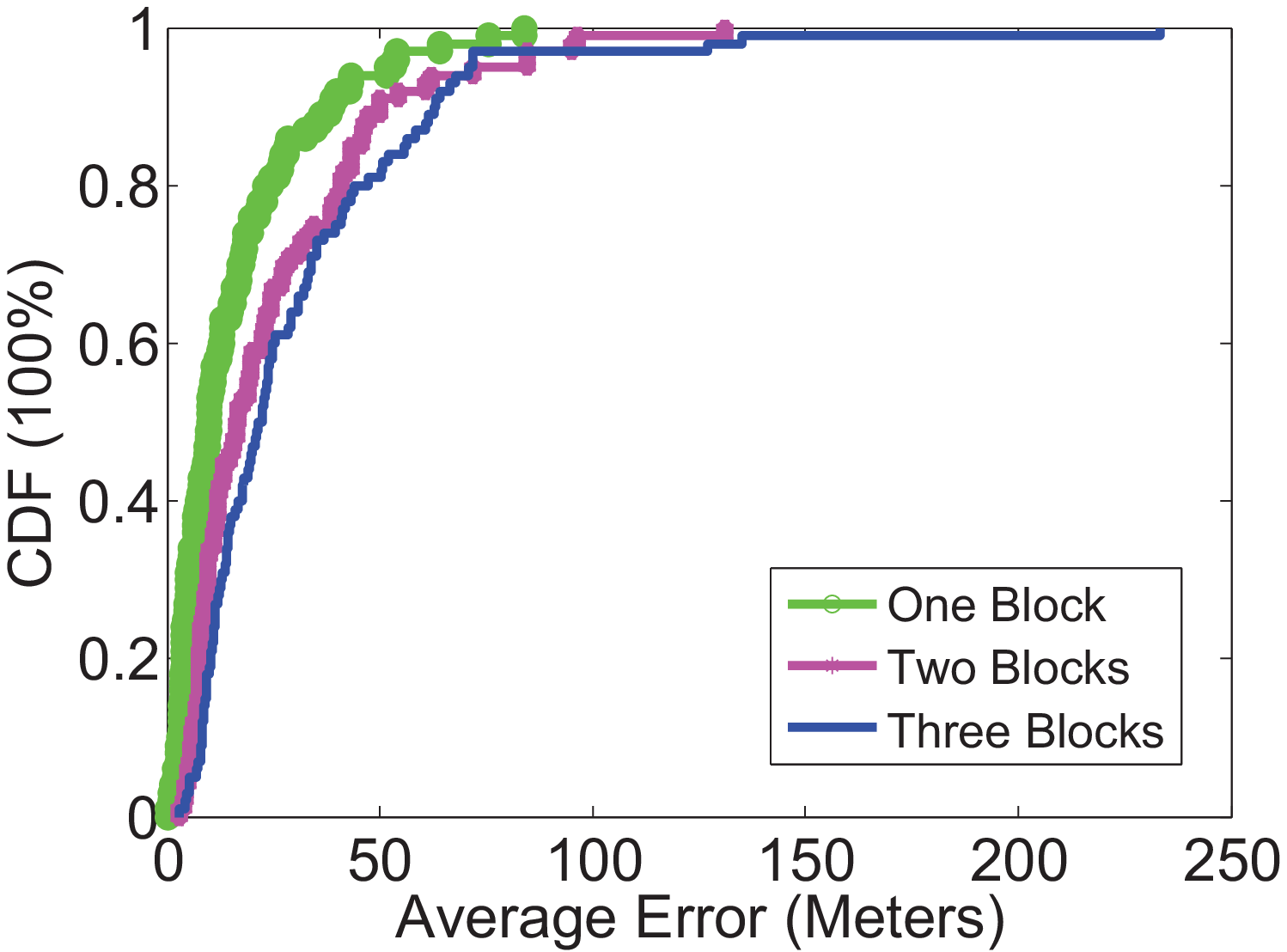}}
        \vspace{-0.1in}
        \caption{Navigation performance evaluation.}
        \vspace{-0.15in}
\label{fig:local_compare3}
\vspace{-0.15in}
\end{figure}
We then  present the localization results in Chicago downtown.
As aforementioned, it is difficult to get the ground
 truth for the majority of the sampling locations.

We set the experiments of estimating the final location. Since,
Section~\ref{sec:preliminary} has demonstrated that there are $9$ bad road
segments with lengths over $400$m,
which is less than $3$ blocks in downtown Chicago. The goal of
\emph{SmartLoc} is then to obtain a relatively  accurate distance
estimation within three blocks.
We randomly select $100$ points as destinations in the experiment, and
a destination could be one block, two
blocks, or three blocks away from the starting point. We drive to
these destination points to evaluate if the destination is precisely
calculated by \emph{SmartLoc}. We assume that the GPS
signals are good before the starting point, and \emph{SmartLoc} will
train the dead-reckoning model during the driving. In
this experiment, we test the accuracy of estimating the traveling
distance in  every time slot and of estimating the overall driving
distance (\ie, locating the final destination) as shown in
Figure~\ref{fig:total_block_seg_cdf} and
Figure~\ref{fig:total_block_cdf} respectively. When \emph{SmartLoc} only
navigates to the destination within one block, with probability $70\%$,
the error of estimating
the location for each sampling slot is less than $10$m, and with
probability $85\%$, the  mean error is less than $30$m.
When the destination is two blocks away,  about $75\%$ of the errors are
less than $30$m;
 when the destination is three blocks away,  about $80\%$
 errors are less than $50$m.
From these figures,  the error of destination  locating
 within a few blocks is acceptable.
We also plot the localization results for one road segment with length
 over $6400$m in Figure~\ref{fig:michi_spot}.
In this figure, the red
 spots denote the ground truth generated from GPS, and the blue
 spots represent the  localization calculated by \emph{SmartLoc}, where the
 green line between them is the localization error for every location.

\section{Conclusion}
\label{sec:conclusion}
This paper presented  \emph{SmartLoc}, a metropolis
 localization system by using the inertial sensors and the GPS module of smartphones.
We  established a predictive regression model to estimate
 the  trajectory  using linear regression, and the proposed \emph{SmartLoc} detects the road infrastructures and
 driving patterns as landmarks to calibrate the localization results.
Our extensive evaluations shows that SmartLoc improves the
 localization accuracy to less than $20m$ for more than $90\%$ roads in
 Chicago downtown, compared with $\geq 50\%$ with raw GPS data.

\small{
%\bibliographystyle{acm}
%\bibliography{reference}

}

\end{document}